\documentclass[aps,pra,reprint,groupedaddress,amsmath,amssymb]{revtex4-1}
\usepackage{bm}
\usepackage{graphicx}
\usepackage[usenames, dvipsnames]{color}
\usepackage[justification=raggedright,singlelinecheck=false]{subcaption}
\usepackage{dsfont}                      
\usepackage[breaklinks,linktocpage]{hyperref}
\graphicspath{{figures/}}

\newcommand*\diff{\mathop{}\!\mathrm{d}}

\newcommand*\smrn{\sum_{|s| \leq 1}\sum_{ |n| \leq \mathcal{N}}}

\begin{document}
\title{A model for calorimetric measurements in an open quantum system}%

\author{Brecht Donvil}%
\email{brecht.donvil@helsinki.fi}
\affiliation{Department of Mathematics and Statistics, University of Helsinki, P.O. Box 68, 00014 Helsinki, Finland}
\author{Paolo Muratore-Ginanneschi}%
\email{paolo.muratore-ginanneschi@helsinki.fi}
\affiliation{Department of Mathematics and Statistics, University of Helsinki, P.O. Box 68, 00014 Helsinki, Finland}
\author{Jukka P. Pekola}%
\email{jukka.pekola@aalto.fi}
\affiliation{QTF Centre of Excellence, Aalto University School of Science, P.O. Box 13500, 00076 Aalto, Finland}
\author{Kay Schwieger}%
\email{kay.schwieger@gmail.com}
\affiliation{iteratec GmbH, Zettachring 6 70567 Stuttgart, Germany}
\date{December, 2017}
\begin{abstract}

We  investigate the  experimental setup  proposed in [New J.   Phys.,
\textbf{15},   115006  (2013)]   for   calorimetric  measurements   of
thermodynamic indicators  in an  open quantum system.   As theoretical
model we consider a periodically driven qubit coupled with a large yet
finite  electron  reservoir,  the  calorimeter.   The  calorimeter  is
initially  at  equilibrium with  an  infinite  phonon bath.   As  time
elapses, the temperature  of the calorimeter varies  in consequence of
energy exchanges  with the qubit  and the  phonon bath.  We  show how
under   weak    coupling   assumptions,    the   evolution    of   the
qubit-calorimeter  system can  be described  by a  generalized quantum
jump process  including as dynamical  variable the temperature  of the
calorimeter.   We  study the  jump  process  by numeric  and  analytic
methods.   Asymptotically with the duration  of   the  drive,   the
qubit-calorimeter  attains a  steady  state.  In  this same limit, we  use
multiscale perturbation  theory to  derive a Fokker-Planck equation governing 
the calorimeter  temperature distribution.   We  inquire the properties of
the temperature probability distribution close and at the steady state.
In particular, we predict the behavior of measurable statistical indicators
versus the qubit-calorimeter coupling constant.

\end{abstract}
\maketitle
\section{Introduction}  

The measurement of thermodynamic quantities  in an open quantum system
poses considerable  experimental challenges.  The main  reason is that
one needs to find  a way to monitor all the  active degrees of freedom
in  the  system  and  its  environment.  

The  proposal  of \cite{PeSoShAv13}  is  to  detect quanta  of  energy
absorbed  or  emitted  by a  driven  quantum  system  by  measuring  the
temperature  variation  of  the   environment  surrounding  it.   More
precisely,  \cite{PeSoShAv13} considers  an integrated quantum circuit
including  a   superconducting  qubit  and  a   resistor  element.   A
superconducting qubit is a two  level artificial atom constructed from
collective  electrodynamic  modes  of  a  macroscopic  superconducting
element  \cite{But87,BoViJoEsDe98}.   Superconducting  qubits  can  be
coupled with other linear circuit elements like capacitors, inductors,
and transmission lines. This fact renders in principle possible to monitor
energy exchanges of the qubit by constantly monitoring the temperature
of a  resistor element in the  circuit.  Hence, the realization  of the
experiment \cite{PeSoShAv13} essentially  hinges upon the feasibility
of measuring the  temperature of the  calorimeter sufficiently accurate over time scales
shorter  than  the thermal  relaxation  time of  the  qubit.
Recent  developments  of  nano-scale  radio-frequency  thermometry
permit to envisage the accomplishment  of this goal.  Already a decade
ago,  \cite{ScYuCl03} demonstrated  the feasibility  of measuring  the
temperature    of    the    normal    metal    side    of    an    SIN
(Superconductor-Insulator-Normal  metal)  tunnel junction  thermometer
with  a  bandwidths  of   up  to  $100\,\mathrm{MHz}$.   More  recently,
\cite{GaViSaFaArMePe15,ViSuGaSaAnPe15} showed that SIN thermometry can
operate  down to  temperatures of  $100\,\mathrm{mK}$ and  detect a  $10
\,\mathrm{mK}$ temperature spike in  a single-shot measurement.  This is
not yet  sufficient for calorimetric measurements  of single microwave
photons in  superconducting quantum  circuit, but makes the prospect of realizing the experiment \cite{PeSoShAv13}  in the
near future very concrete.

The  aim  of the  present  contribution  is to  theoretically  explore
features of the temperature process  in \cite{PeSoShAv13}.  We take as
starting point  the theoretical qubit-calorimeter model  introduced in
\cite{KuMGPeSc16}.   Accordingly, we  describe the  dynamics of  the
qubit-calorimeter  system  by  a   generalized  quantum  jump  process
\cite{DaCaMo92,Car93}.   The generalization  consists  in treating  as
dynamical variable the \emph{temperature  of the calorimeter} together
with the components  of the state vector of the  qubit. The derivation
of  the quantum  jump  process  then follows  from  the  usual set  of
assumptions presiding  over the validity of the Markovian  approximation (see
for example   \cite{BrPe02}) and  the hypothesis  that in  between
interactions with the qubit the calorimeter  behaves as a Fermi gas in
local equilibrium.   In other words,  the calorimeter is modelled  by a
collection of grand-canonical ensembles  parametrized by a temperature
evolving  in  time  according   to  a  prescribed  dynamics.   Extended
statistical  ensembles  characterized   by  a  dynamically  determined
temperature come naturally about, for example in the study of energy exchanges
between a  single electron  box tunnel  coupled to  metallic reservoir
\cite{BeBrSa15}, and in macroscopic statistical physics for example as
a tool to optimize Monte Carlo methods \cite{MaPa92}.

As a  step towards  increased realism, we advance the model of \cite{KuMGPeSc16}
in two ways. First, we suppose that the  qubit  is strongly coupled 
with a periodic  control  field.  Drawing  on
\cite{BrPe97},  we obtain  the corresponding  stochastic Schr\"odinger
equation for the qubit. Second, we include in the model normal metal 
electron-phonon interactions between the calorimeter and the environment. 
Electron-phonon interactions bring about  a drift and a  noise term 
in the  stochastic differential equations governing the calorimeter 
temperature \cite{KaLiTa57,WeUrCl94,PeKa18}.

We  then inquire  the  behaviour of the  probability  distribution of
temperature $T_e$ of  the calorimeter by numeric  and analytic methods
and for experimentally relevant values of the parameters. We show that
as the duration  of the drive increases  the temperature distribution
tends to an equilibrium state.  In order to shed more light on the asymptotic
stage of  the dynamics,  we  take  advantage  of the  time-scale
separation between  the characteristic  relaxation times of  the qubit
and the  temperature process and show by means of multiscale perturbation theory \cite{PaSt08} 
that  the temperature probability distribution evolves
asymptotically according  to a Fokker-Planck  equation \cite{Gardiner}. 
The Fokker--Planck equation evinces the general form of dependence upon the phonon temperature $T_p$ and the qubit-calorimeter coupling $g$ of the steady state temperature $T_S$ and the temperature distribution relaxation time to equilibrium $\tau_S$.

The    structure     of    the    paper    is     as    follows.    In
section~\ref{sec:experiment} we briefly sketch the
experimental setup  of \cite{PeSoShAv13}.   In section~\ref{sec:model}
we introduce the qubit-calorimeter model whose dynamics in the Markovian limit 
we subsequently present in  section~\ref{sec:quT}.   In section  \ref{sec:master}  we
inquire  the  asymptotic  behavior   of  the  temperature  probability
distribution by  multiscale methods.   Finally, we report on our numeric
investigation of the model in  section~\ref{sec:sim}.  We focus on two
regimes. The  first regime  or "short  time regime"  is 10  periods of
resonant frequency. In this time the qubit and temperature only make a
few jumps. The second regime or the "long time" regime is of the order
of $10^4$ periods of resonant frequency. On this time scale the system
makes many jumps and drift term  due to the phonons becomes important.
In the physically relevant parametric range, the  results of  the 
simulations are in good agreement  with the  analytic predictions of 
section \ref{sec:master}.

Finally,  we defer most of the technical calculations to the appendices.

\section{The Qubit-Calorimeter circuit}
\label{sec:experiment}

Superconducting qubits  are solid state devices  behaving according to
the   rules  of   quantum  mechanics.    They  combine   the  feature,
characteristic of  atoms, of  exhibiting quantized energy  levels with
the flexibility of  linear circuit elements which can  be connected in
more  complex networks.  The  realization of  a superconducting  qubit
plays   upon   the   properties    of   Josephson   tunnel   junctions
\cite{But87,Leg92}.   Namely,  at  temperatures  sufficiently  low  to
render thermal  noise negligible, Josephson tunnel  junctions maintain
quantum coherence  of  charge transport (i.e.   are non-dissipative)
governed     by    a     non-harmonic     Hamiltonian    (see     e.g.
\cite{BoViJoEsDe98,DeWaMa03,DeSc13}).   Non-linear  separation of  the
energy levels, is essential to  prevent qubit operations from exciting
transitions between more than two states in the system.

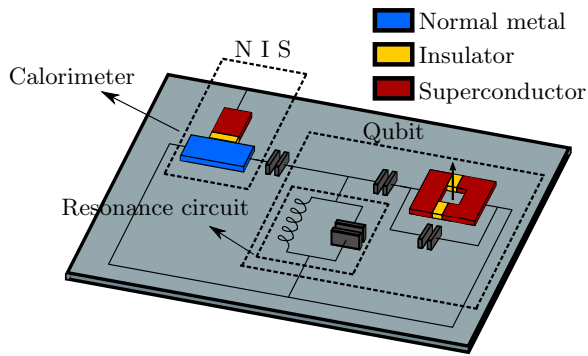
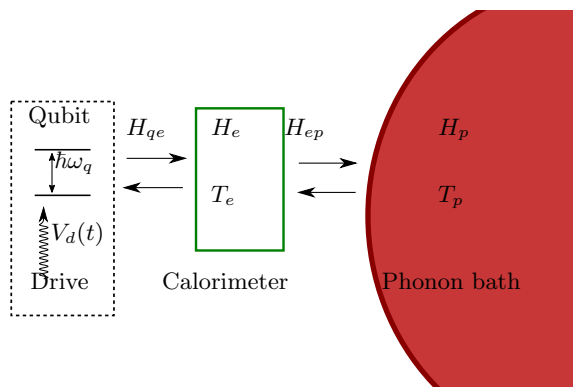
\begin{figure}
\centering
\begin{subfigure}[t]{0.47\textwidth}
\centering
\includegraphics[scale=0.15]{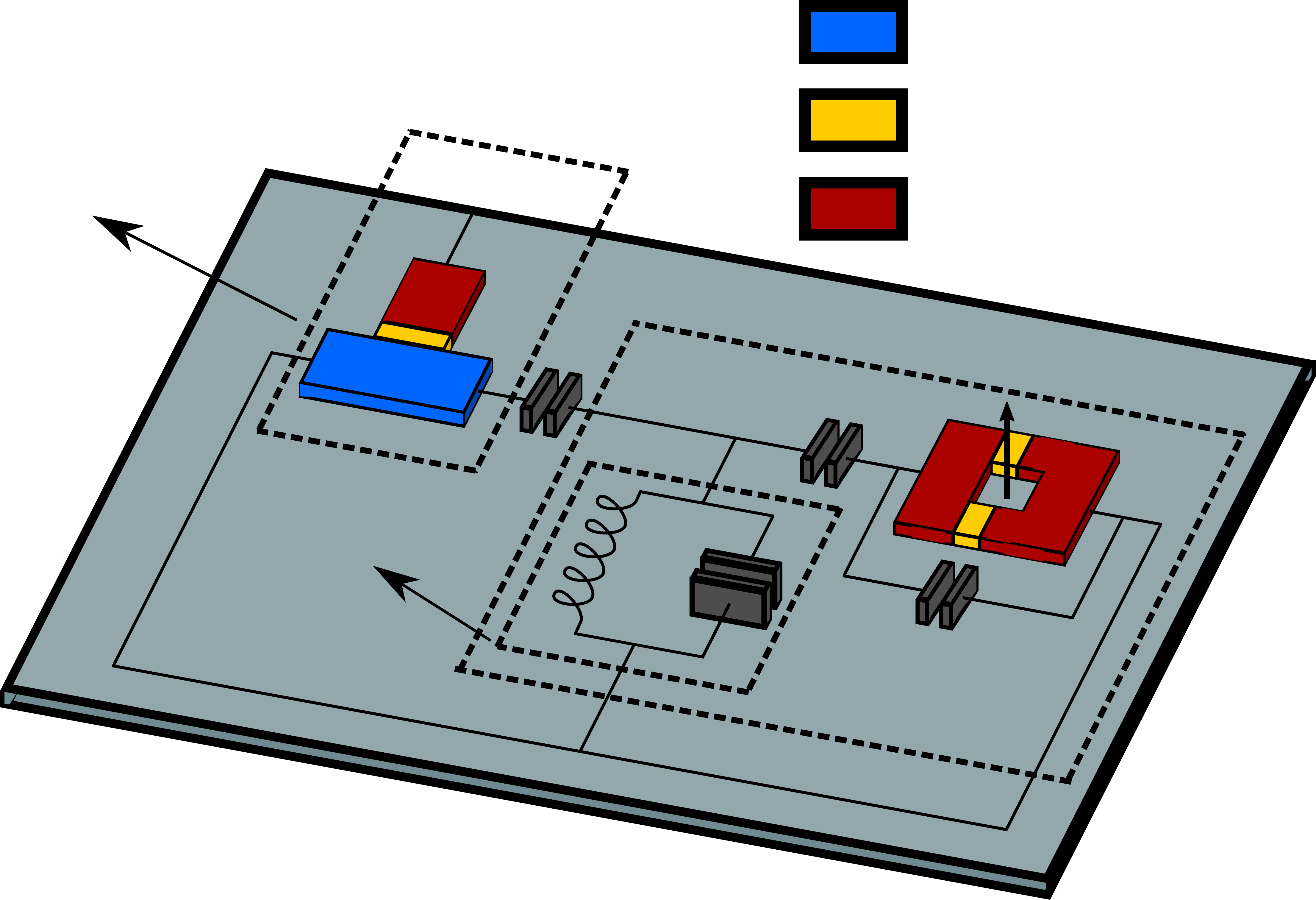}
\setlength{\unitlength}{0.1cm} 
\begin{picture}(0,0)
\put(-47,39.5){N I S}
\put(-70,18.5){Resonance circuit}
\put(-77,36){Calorimeter}
\put(-30,28.7){Qubit}
\put(-22.5,43.3){Normal metal}
\put(-22.5,38.8){Insulator}
\put(-22.5,34.0){Superconductor}
\end{picture}
\caption{The   quantum   integrated  circuit of
\cite{PeSoShAv13}.  The temperature measurement  is performed by embedding
an N I S junction in a resonance circuit. The calorimeter consists of the
 electrons in the normal metal.  The transmon qubit is formed
by a Cooper pair box embedded in a resonance circuit. This figure is
not up to scale. The Cooper pair box is of the order of 10 $\mu$m, the
resonance circuit which it is embedded in is of the order of 1 mm, and
the calorimeter is 1 $\mu$m.}
\label{fig:exp}
\end{subfigure}
\begin{subfigure}[t]{0.47\textwidth}
\centering
\includegraphics[scale=0.25]{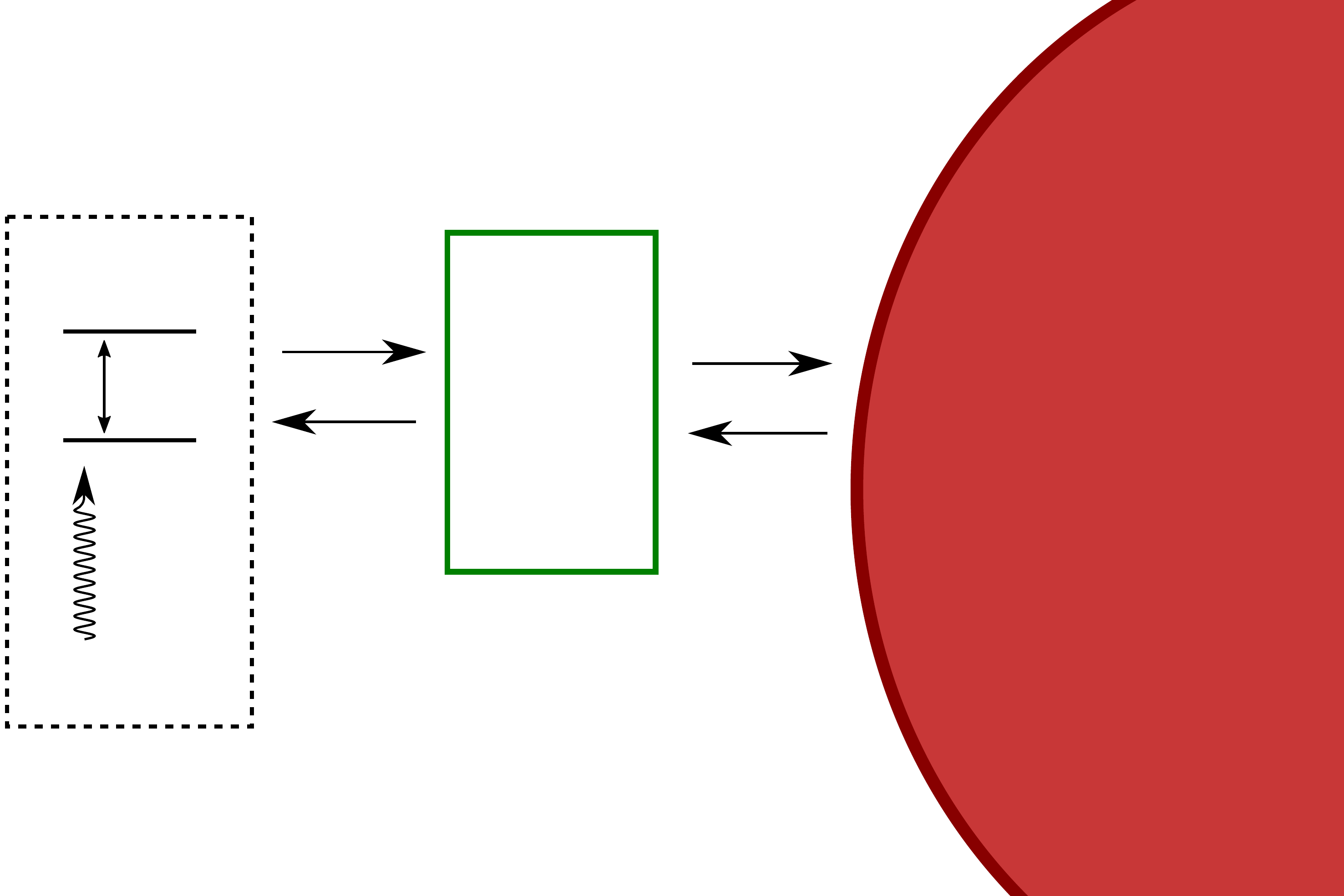}
\setlength{\unitlength}{0.1cm} 
\begin{picture}(0,0)
\put(-71,29){$\hbar\omega_q$}
\put(-71.5,19.5){$V_d(t)$}
\put(-74.2,13){Drive}
\put(-74.5,35){Qubit}
\put(-61.5,33.5){$H_{q e}$}
\put(-56.7,13){Calorimeter}
\put(-50.2,33.5){$H_{e}$}
\put(-50.2,24.5){$T_{e}$}
\put(-40.5,33.5){$H_{e p}$}
\put(-20,33.5){$H_{p}$}
\put(-20,24.5){$T_{p}$}
\put(-27.5,13){Phonon bath}
\end{picture}
\caption{The qubit-calorimeter experiment as modelled by the Hamiltonian \eqref{model:H}.}
\label{fig:model}
\end{subfigure}
\caption{Visual representation of the   quantum   integrated  circuit 
and its mathematical model.
}
\end{figure}

In Figure~(\ref{fig:exp}) we draw a  quantum integrated circuit of the
type envisaged in \cite{PeSoShAv13}.   The circuit contains a transmon
qubit   \cite{KoYuGaHoSc07}.   A transmon   qubit consists  of  a
superconducting island coupled through Josephson junctions and shunted
by a capacitor. The transmon qubit  is embedded in a resonance circuit
to amplify its signal. The  resistor element in the  circuit (bright
blue  online in  figure~(\ref{fig:exp}))  is the  calorimeter, making it into an open quantum system. 
In the current work we do not consider the resonator, but model the qubit to be directly coupled to the calorimeter. We 
conceptualize the calorimeter as a gas of free electrons weakly interacting with an
infinite  phonon  thermal bath.  Phonons describe excitations of the 
lattice structure of the normal metal in the resistor and in the circuit substrate. 
The phonon bath is maintained at a uniform constant temperature $T_{p}$ equal to 
that of the cryostat. The temperature  $T_{e}$ of  the electron  gas is  in  equilibrium with the phonon bath at the beginning of the experiment.  The
drive is an external  periodic control potential initially turned-off.
When  turned on,  the drive  excites  transitions in  the qubit.   The
temperature of the  resistor varies then in consequence  of the single
microwave photons  emitted or absorbed  by the two-level  system.  The
actual    temperature    measurement     happens    via    a    Normal
metal-Insulator-Superconductor              (N I S)             junction
\cite{ScYuCl03,GaViSaFaArMePe15,ViSuGaSaAnPe15} on  the resistor.  This is
possible because the conductance $G$ of the N I S junction   depends on the
temperature $T_e$ of  the normal metal whereas it is  \emph{independent} 
of the temperature of the superconductor:
\begin{equation}
G= \int_{\mathbb{R}}\frac{\diff E \, N_S(E)}{R_T\,k_{\mathrm{B}}\,T_e}f_{T_{e}}(E-e\,V_b)\big{[}1-f_{T_{e}}(E-e\,V_b)\big{]}.
\nonumber
\end{equation}
Here,  $N_S(E)$ is the normalized Bardeen-Cooper-Schrieffer superconducting 
density of states, $k_{\mathrm{B}}$ is Boltzmann constant and
$f_{T_{e}}(E)=(1+\exp(E/k_{\mathrm{B}}T_e))^{-1}$ the Fermi--Dirac distribution at temperature $T_{e}$ with 
$E$ referenced to the chemical potential \cite{GiHeLuSaPe06}. 
$V_{b}$ and $R_{T}$ are respectively the voltage bias and the resistance of the 
N I S junction.

\section{Theoretical Model}  
\label{sec:model}  

Figure \ref{fig:model} graphically  illustrates our mathematical model
of  the  qubit-calorimeter-phonon   interactions.   The
Schr\"{o}dinger picture Hamiltonian of the  full quantum system is 
the sum
\begin{eqnarray}
H=H_{q}+H_{q e}+H_{e}+H_{e p}+H_{p}
\label{model:H}
\end{eqnarray}
of the qubit $H_{q}$, the qubit-electron interaction $H_{q e}$,
the electron gas $H_{e}$, the electron-phonon interaction $H_{e p}$ and the
phonon $H_{p}$ Hamiltonians.

The Hamiltonian of the qubit is 
\begin{equation} 
H_q(t)=\frac{\hbar\,\omega_q}{2}\sigma_z +\kappa\,V_{d}(t)
\label{model:qubit}
\end{equation}  
where $\sigma_z$  denotes the  diagonal Pauli  matrix.
The Hamiltonian is  time non-autonomous owing to the presence of the driving
potential   $V_{d}(t)$.  The   drive   is  periodic  with   frequency
$\omega_{L}$ and is strongly coupled  with the qubit by the non-dimensional parameter $\kappa$.  
In consequence, it is expedient to resort to Floquet theory to describe the periodically 
driven qubit dynamics \cite{Shi65,Zel67,Sam73} see also \cite{GrHa98,Hol15,Don18} and 
appendix~\ref{ap:Floquet}.

The qubit is directly coupled only to the calorimeter via the Hamiltonian
\begin{equation}
\label{model:hInt}
H_{q e}=g\,\frac{\sqrt{8\,\pi}\,\epsilon_{F}}{3\,N}
\sum_{k\,\neq\,l\,\in\,\mathbb{S}}(\sigma_{+}+\sigma_{-})\,a^{\dagger}_{k}\,a_{l}
\end{equation}  
Here  $\sigma_{+}$  and  $\sigma_{-}$  are  the  qubit
raising  and  lowering operators  in  the  absence of  external  drive
$V_{d}$. Similarly, $a_{k}$ and $a_{k}^{\dagger}$ are the annihilation
and creation operators of a free  fermion with energy specified by the
absolute value  $k$ of its  wave number $\boldsymbol{k}$.  The  sum in
(\ref{model:hInt}) is restricted to an energy shell $\mathbb{S}$ close
to the Fermi  energy $\epsilon_{F}$ of the metal in  the resistor. The
sum ranges over non-diagonal terms to avoid trivial renormalization of
the  energy levels  of  the non-interacting  Hamiltonians $H_{q}$  and
$H_{e}$. We  choose the numerical prefactor  in (\ref{model:hInt}) for
computational  convenience.   The  interaction   is  strength is characterised  by  the
non-dimensional  constant  $g\,\ll\,1$,  and $\epsilon_{F}$  sets  the
energy scale. Finally, $N=O(10^{9})$ is the number of electrons in the
shell $\mathbb{S}$.

Of  the  remaining three  terms  on  the  right hand  side  of
(\ref{model:H}), $H_{e}$  and $H_{p}$ are  the free fermion  and boson
gases Hamiltonians  weakly coupled  by a Fr\"{o}lich  interaction term
$H_{e p}$ \cite{Fro52}  (see for example chapter~9  of \cite{Mah11}). As
these  Hamiltonians   are  textbook   knowledge,  we   defer  explicit
expressions  and  quantitative analysis  to  appendix~\ref{sec:appPh}.
Here we discuss the qualitative picture.  Phonons describe
small  vibrations  in the  lattice  structure  of  the metal  and  its
substrate.  Phonon-self interactions can be neglected as the vibration
amplitude is  small with  respect to the  characteristic length  of the
lattice  cell  $O(k_{F}^{-1})$  for  $k_{F}$  the  Fermi  wave  vector
\cite{AsMe76}.  Electron  self-interactions  are  re-absorbed  in  the
parameters  of   the  free  energy  spectrum.    Namely,  the  typical
relaxation  rate to  the  Fermi--Dirac energy  distribution of  Landau
quasi-particle  in   a  metallic  wire  is   of  the  order  of   $\tau_{e e}\sim 1$ ns
\cite{PoGuBiNoEsDe97}, whereas  electron-phonon interactions typically
occur on a $\tau_{e p}\sim 10^4$ ns  time-scale \cite{GaViSaFaArMePe15}. Thus, at any
instant of  time the state  of phonons  and electrons is  described by
quantum statistical equilibrium ensembles at well defined temperatures
respectively  denoted by  $T_{p}$ and  $T_{e}$.  Within  leading order
accuracy,  equilibrium states  are perturbed  by the  deformation term
$H_{e p}$.  In  the presence of small differences  between $T_{p}$ and
$T_{e}$, the perturbation  results in a mean  energy current $J\propto
T_{p}^{5}-T_{e}^{5}$  
\cite{KaLiTa57,WeUrCl94} with  root mean  square
fluctuations $O(T_{p}^{3})$ \cite{PeKa18}.   Experiments at sub-Kelvin
temperatures clearly support these theoretical estimates.

\section{Qubit-Calorimeter Process}
\label{sec:quT}

Typical  transmon  qubit   relaxation  times  are  of   the  order  of
$\tau_{R}\sim 2-5\times  10^5$ ns  \cite{WaAxDeSc15}.  The  time scale
separation $\tau_{e  e}/\tau_{R}\sim 10^{-5}$, suggests  describing
the qubit  dynamics in the Born--Markov  approximation.  The Markovian
approximation  is consistent  with  complete positivity  of the  state
operator  if  we  retain  only  secular terms  in  the  evaluation  of
transition  rates  \cite{BrPe02}.   The  rotating  wave  approximation
offers a  systematic procedure  to neglect  non-secular terms.   It is
justified    if   transitions    among   quasi-energy    levels   (see
appendix~\ref{ap:Floquet}) in the qubit occur with rates much smaller
than  the  corresponding  frequency  gaps in  the  radiation  spectrum
emitted by the qubit \cite{BrPe97,BrPe02}.  In other words, we need to
work under the hypothesis that  characteristic time scale $\tau_{q e}$
of  the qubit-calorimeter  interaction is  much larger  than the  time
$\tau_{m}$  set by  the typical  inverse  separation of  peaks in  the
radiation spectrum.   In the weak-coupling limit,  Fermi's golden rule
self-consistently  yields the  estimate $\tau_{e  q}\sim g^{-2}$.   We
then expect the Markovian approximation to hold in the presence of
a  strongly  coupled  drive in  (\ref{model:qubit})  
if  $\tau_{m}\sim
\kappa^{-1}$ holds so  that $\tau_{m}/\tau_{e q}\sim g^{2}/\kappa\ll
1$.   We  verify  this  assumption  
in  section~\ref{sec:sim}  for  an
explicit, experimentally  relevant drive.  Finally, the  evaluation of
qubit  transition rates  using  the  Fermi-Dirac distribution  imposes
$\tau_{e q}\,\ll\,\tau_{e p}\,\ll\,\tau_{R}$.

Under the above assumptions \cite{KuMGPeSc16}, we unravel the Markovian approximation 
for the qubit dynamics in the form of a Poisson-stochastic Schr\"odinger equation 
\cite{BrPe97,BrPe02}
\begin{eqnarray}
\label{eq:sse}
\lefteqn{
\diff \psi(t) = -\frac{\imath}{\hbar} \,G(\psi(t))\diff t
}
\nonumber\\&&
+\smrn\left(\frac{A_{s,n}\psi(t)}{\|A_{s,n}\psi(t)\|}
-\psi(t)\right)\diff \nu_{s,n}(t)
\end{eqnarray}
The vector $\psi\in\mathbb{C}^{2}$ instantaneously specifies the state of the qubit. 
The sums on the right hand side ranges over 
the Lindblad operators
\begin{eqnarray}
\label{eq:jumpop1}
\lefteqn{A_{s,n}=
}
\nonumber\\&&
\delta_{s,0}\,D_{1,1,n}\,\Big{(}|\phi_{1,0}(0)\rangle\langle \phi_{1,0}(0)|
-|\phi_{0,0}(0)\rangle\langle \phi_{0,0}(0)|\Big{)}
\nonumber\\&&
+(\delta_{s,1}+\delta_{s,-1})\,D_{\frac{1+s}{2},\frac{1-s}{2},n}\,
\Big{|}\phi_{\frac{1+s}{2},0}(0)\Big{\rangle}\Big{\langle} \phi_{\frac{1-s}{2},0}(0)\Big{|}.
\end{eqnarray}
By $\phi_{r,n}$ $r=0,1$, $n\in\mathbb{Z}$ we denote the element of the
orthonormal basis in $\mathbb{C}^{2}\times\mathbb{L}^{2}[0,2\,\pi/\omega_{L}]$
associated to the quasi-energy level
\begin{eqnarray}
\epsilon_{r,n}=\epsilon_{r}+n\,\hbar\,\omega_{L}
\label{quT:quasi}
\end{eqnarray}
specified by Floquet theory (see appendix~\ref{ap:Floquet} and references therein). 
For any fixed $t$, the vector functions $\phi_{r,0}(t)$, $r=0,1$ form 
an orthonormal basis with respect to the standard scalar product in $\mathbb{C}^{2}$. 
The selection rules imposed by the matrix elements
\begin{eqnarray}
\label{eq:jumpop2}
D_{r,r^{\prime},n}=\frac{2\,\pi}{\omega_{L}}\int_{0}^{\frac{2\,\pi}{\omega_{L}}}
\mathrm{d}t\,\langle\phi_{r,n}(t)|\sigma_{+}+\sigma_{-}|\phi_{r^{\prime},0}(t)\rangle
\end{eqnarray}
restrict the number of non-vanishing Lindblad operators (\cite{BrPe97,BrPe02} 
and appendix~\ref{ap:qe}). The representation (\ref{eq:jumpop1}) holds under the 
simplifying but not too restrictive assumption that there is a one to one 
correspondence between  Lindblad operators and frequencies in the qubit radiation 
spectrum
\begin{eqnarray}
\omega_{s,n}=s\,\frac{\epsilon_{0}-\epsilon_{1}}{\hbar}-n\,\omega_{L}
\label{eq:selection}
\end{eqnarray}
for $s=0,\pm 1$ and $n=0,\pm 1,\dots,\pm \mathcal{N}$. 
This  assumption is reasonable if the selection rule (\ref{eq:jumpop2})
yields non-vanishing contribution only for a finite number of transitions, i.e. $\mathcal{N}<\infty$.
In particular, it holds true for monochromatic drive we consider 
in  section~\ref{sec:sim}.

The evolution law (\ref{eq:sse}) specifies a piecewise deterministic process. 
The deterministic evolution corresponds to a first order differential equation 
in $\mathbb{C}^{2}$ governed by the non-linear norm preserving drift
\begin{eqnarray}
\label{eq:contev}
\lefteqn{
G(\psi)=
}
\nonumber\\&&
\frac{\imath\,\hbar}{2}
\smrn\Gamma(\omega_{s,n},T_{e})(\|A_{s,n}\psi\|^2- A^\dagger_{s,n}A_{s,n})\psi
\end{eqnarray}  
where $\|\cdot\|^{2}=\langle\cdot|\cdot\rangle$ is the squared norm in $\mathbb{C}^{2}$.
The deterministic evolution is interrupted at random times by jumps modeled by
the increment $\diff \nu_{s,n}(t)$ of statistically independent 
Poisson processes for each $s,n$ and fully characterized by the conditional 
expectation
\begin{equation}\label{eq:condav}
\mathbb{E}(\diff \nu_{s,n}(t)|\psi)=\Gamma(\omega_{s,n},T_{e})\|A_{s,n}\psi\|^2\diff t.
\end{equation}
Here and in (\ref{eq:contev}) the radiation frequency dependence of
\begin{equation}
\label{eq:gam}
\Gamma(\omega,T_{e})=\frac{g^2\,\omega\, e^{\hbar\omega/(k_{\mathrm{B}}T_e)}}{e^{\hbar\omega/(k_{B}\,T_e)}-1}.
\end{equation} 
stems from the fact that a leading order transitions in the qubit always involves creating 
and annihilating an electron in the calorimeter (see appendix~\ref{ap:qe} for details).
For $\omega>0$ the calorimeter absorbs energy, for $\omega<0$ the calorimeter 
loses energy. The  rates depend  on the  temperature of  the electron bath $T_e$.

We determine the temperature $T_{e}$ from internal energy $E$ of the calorimeter  
using the  Sommerfeld approximation, see e.g.  \cite{AsMe76}.  
Under our working assumptions, $\mathrm{d}E$ is non-vanishing
only over time-scales larger than $\tau_{e q}$. We obtain
\begin{equation}\label{eq:temptoE}
\diff T_e^2(t)=\frac{1}{N\,\gamma}\diff E(t),
\end{equation}
where
\begin{equation}\label{eq:heatcap}
\gamma=\frac{\pi^2\,k_B^2}{4\,\epsilon_F}.
\end{equation}  
According to these definitions  $N\,\gamma/2$  is  the  coefficient  of  the  linear
contribution to the heat capacity.

We identify two main contributions to the right hand side of (\ref{eq:temptoE}):
\begin{eqnarray}
\mathrm{d}E(t)=\mathrm{d}E_{e q}(t)+\mathrm{d}E_{e p}(t)
\nonumber
\end{eqnarray}
A jump in the qubit donates $\hbar\omega_{s,n}$ to
the calorimeter.  The corresponding instantaneous change in energy is
\begin{eqnarray}
\mathrm{d}E_{e q}(t)=\smrn\hbar\omega_{s,n}\diff  \nu_{s,n}(t)
\label{eq:deeq}
\end{eqnarray}
The  increment $\mathrm{d}E_{e p}$ embodies the contribution of  electron-phonon 
interactions. We model these interactions as the sum of a deterministic 
and a stochastic differential \cite{KaLiTa57,WeUrCl94,PeKa18}
\begin{eqnarray}
\hspace{-0.5cm}\mathrm{d}E_{e p}(t)=\Sigma V(T_p^{5}-T_e^{5}(t))\diff t+\sqrt{10\Sigma V k_B}T_p^3 \diff w(t).
\label{eq:deep}
\end{eqnarray}
Here $\mathrm{d}w(t)$ is the increment of a one-dimensional Wiener process, 
 $\Sigma$ is a  material constant defined in Appendix~\ref{sec:appPh}, $V$ is the volume of the calorimeter 
and $T_p$ is the temperature of the phonon bath. The drift term in (\ref{eq:deep})
tends to bring back the calorimeter into equilibrium with the phonon bath
a temperature $T_{p}$. The Wiener increments models fluctuation of the
heat current between the calorimeter and the phonon reservoir. Within leading 
accuracy, we evaluate the characteristic size of heat fluctuations by setting $T_{p}=T_{e}$.
The diffusion coefficient in (\ref{eq:deep})
does not prevent by construction realizations of the temperature process from acquiring 
nonphysical negative values. This means that (\ref{eq:temptoE}) must be complemented by
proper, e.g. reflecting, boundary conditions at $T_{e}=0$. Physically, the barrier at vanishing 
temperature can be understood observing that the energy distribution of 
a finite sized free-electron reservoir vanishes at low energies with a sharp drop 
to zero at the energy corresponding to the filled Fermi sea \cite{PeMGKuGa16}.

We are mainly interested  in the
evolution of the calorimeter temperature $T_{e}$. However,  in order to  obtain numerical
results, it is necessary to  simulate both processes: the evolution of
the  qubit \eqref{eq:sse}  and  the  temperature \eqref{eq:temptoE}  are
coupled by (\ref{eq:deeq}), (\ref{eq:deep}).  Furthermore, the jump  rates of  
the qubit  \eqref{eq:gam} depend  on the
current  temperature  $T_e$ and on the current state of the qubit by equation
\eqref{eq:condav}. Quantitative predictions about evolution of the
qubit-calorimeter system call for numeric investigation.
It is, however, remarkable that in the long time limit, it
is possible to derive a closed Fokker-Planck equation for the calorimeter 
temperature distribution, as we show in the next section.

\section{Effective Temperature Process}
\label{sec:master}

To start with, it is expedient to define the process $\xi(t)=T_e^2(t)$ which by \eqref{eq:temptoE} 
(\ref{eq:deeq}), (\ref{eq:deep}) obeys the Wiener-Poisson stochastic differential equation
\begin{eqnarray}
\lefteqn{
\mathrm{d}\xi(t)=
\smrn\frac{\hbar\omega_{s,n}}{N\,\gamma}\diff \nu_{s,n}(t)
}
\nonumber\\&&
+\frac{\Sigma V(T_p^{5}-\xi^{5/2}(t))\diff t+\sqrt{10\Sigma V k_B}T_p^3 \diff w(t)}{N\,\gamma}.
\label{master:xi}
\end{eqnarray}
In  Appendix \ref{sec:apMaster} we show that the joint 
probability 
\begin{eqnarray}
\lefteqn{\hspace{-0.4cm}
P_{r}(X,t)=
}
\nonumber\\&&\hspace{-0.4cm}
\mathrm{P}\Big{(}X\leq \xi(t)<X+\mathrm{d}X \,\&\,\mbox{qubit in Floquet state}\,r\Big{)}
\label{master:prob}
\end{eqnarray}
defined by (\ref{eq:sse}), 
(\ref{master:xi}) obeys a closed time-autonomous 
Chapman--Kolmogorov master equation
\begin{eqnarray}
\label{master:master}
\lefteqn{
\dot{P}_{r}(X,t)=\mathcal{L}_{X}P_{r}(X,t)
}
\nonumber\\&&
+\sum_{r^{\prime}=0,1}\int_{0}^{\infty}\mathrm{d}Y\, 
K_{r r^{\prime}}(X|Y)\,P_{r^{\prime}}(Y,t)
\nonumber\\&&
-\sum_{r^{\prime}=0,1}\int_{0}^{\infty}\mathrm{d}Y\, 
 K_{r^{\prime} r}(Y|X) P_{r}(X,t).
\end{eqnarray}
The differential operation $\mathcal{L}_{X}$ represents the effect of 
electron-phonon interactions 
\begin{align}
\mathcal{L}_{X}P_{r}(X,t)=&-\frac{\Sigma V}{N\,\gamma}\partial_X\big((T_p^{5}-X^{5/2})
P_{r}(X,t)\big)
\nonumber
\\&+\frac{(\sqrt{10\Sigma V k_p}T_p^3 )^2}{2\,N^{2}\,\gamma^{2}}\partial_X^2P_{r}(X,t).
\label{master:FP}
\end{align} 
The  kernel $K$ describes quantum jumps
\begin{eqnarray}
\lefteqn{
K_{r r^{\prime}}(X|Y)=
}
\nonumber\\&&
\sum_{n\leq|\mathcal{N}|}\mathcal{G}_{r,r^{\prime},n}(Y)\,\delta\left(Y-X+\frac{\hbar\,\omega_{r^{\prime}-r,n}}{N\,\gamma}\right)
\label{master:rate1}
\end{eqnarray}
where
\begin{eqnarray}
\mathcal{G}_{r,r^{\prime},n}(X)=\Gamma(\omega_{r-r^{\prime},n},\sqrt{X})\,|D_{r,r^{\prime},n}|^{2}
\label{master:rate2}
\end{eqnarray}
Chapman--Kolmogorov master equations of the type (\ref{master:master})
are compatible with the existence of an $H$-theorem, see e.g. \S~3.7.3
of \cite{Gardiner}. We expect therefore that in the limit of long duration
of the drive, (\ref{master:FP}) admits a steady state and that solutions
corresponding to physical initial data relax to such steady state.

The occurrence of the non-dimensional weighting prefactor
\begin{eqnarray}
\varepsilon=\frac{1}{N}
\nonumber
\end{eqnarray}
in (\ref{master:FP}) evinces the possibility to apply multiscale perturbation theory
\cite{PaSt08} to the asymptotic analysis of the master equation (\ref{master:master}).
Namely, we expect temperature equilibration to occur on a much longer time scale compared to 
the characteristic qubit relaxation time. Formally, if we posit
\begin{eqnarray}
P_{r}(X,t)\equiv \tilde{P}_{r}(X,t,\varepsilon\,t)
\nonumber
\end{eqnarray}
we can couch the time derivative of the probability in terms of the sum
of partial derivatives
\begin{equation}
\frac{\diff P_{r}}{\diff t}(X,t)=
\partial_{ t}\tilde{P}_{r}(X,t,\tau)+\varepsilon\,\partial_{\tau} \tilde{P}_{r}(X,t, \tau)
\end{equation}   
with respect to the ``fast'' variable $t$ and the ``slow'' one
\begin{eqnarray}
\tau=\varepsilon\,t.
\nonumber
\end{eqnarray}
In the limit of long duration of the drive it is then reasonable to assume
that the probability becomes stationary with respect to the fast time dependence
\begin{eqnarray}
\partial_{t}\tilde{P}_{r}=0.
\nonumber
\end{eqnarray}
Therefore, under our working assumption
\begin{eqnarray}
\bar{P}_{r}(X,\tau)\equiv\lim_{t\uparrow\infty}\tilde{P}_{r}(X,t,\tau)
\nonumber
\end{eqnarray}
satisfies
\begin{eqnarray}
\lefteqn{
\varepsilon\,\partial_{\tau}\bar{P}_{r}(X,\tau)=
\varepsilon\,
\mathcal{L}_{X}^{(1)}\bar{P}_{r}(X,\tau)+\varepsilon^{2}\,\mathcal{L}_{X}^{(2)}\bar{P}_{r}(X, \tau)
}
\nonumber\\&&
+\sum_{s=0}^{1}\bigg(\mathds{G}^{(0)}_{r,s}(X)\bar{P}_{s}(X,\tau)
-\mathds{G}^{(0)}_{s, r}(X)\bar{P}_{r}(X,\tau)\bigg)
\nonumber\\&&
+\sum_{k}^{\infty}\frac{\varepsilon^{k}}{k!}\sum_{s=0}^{1}
\partial_X^{k}\bigg(\mathds{G}_{r, s}^{(k)}(X)\,\bar{P}_{s}(X,\tau)\bigg)
\label{master:multiscale}
\end{eqnarray}
In (\ref{master:multiscale}) we use the notation
\begin{equation}
\mathcal{L}_{X}^{(1)}\bar{P}_{r}(X,\tau)
=-\frac{\Sigma\,V}{\gamma}\partial_X\big((T_p^{5}-X^{5/2})\bar{P}_{r}(X,\tau)\big)
\nonumber
\end{equation}
\begin{equation}
\mathcal{L}_{X}^{(2)}\bar{P}_{r}(X,\tau)=\frac{(\sqrt{10\,\Sigma\, V\, k_p}T_p^3 )^2}
{2\,\gamma^2}\partial_X^2\bar{P}_{r}(X,\tau)
\nonumber
\end{equation}
and
\begin{eqnarray}
\mathds{G}_{r,s}^{(k)}(X)=
\sum_{|n|\leq\mathcal{N}}\left(\frac{\hbar\,\omega_{s-r,n}}{\gamma}\right)^{k}\mathcal{G}_{r,s,n}(X).
\nonumber
\end{eqnarray}
We look for solutions of (\ref{master:multiscale}) by expanding 
the probability distribution in an Hilbert series in 
powers of $\varepsilon$
\begin{equation}
\bar{P}_{r}(X,\tau)=\sum_{n=0}^{\infty}\varepsilon^n\,\bar{P}^{(n)}_{r}(X,\tau)
\nonumber
\end{equation}
We readily see that the zero order of the expansion is amenable to the
form
\begin{eqnarray}
\bar{P}^{(0)}_{r}(X,\tau)=Q_{r}(X)\,F^{(0)}(X,\tau)
\nonumber
\end{eqnarray}
The quantity $Q_{r}$ ($r=0,1$) is the population of 
the Floquet state $\phi_{r,0}(0)$ at thermal equilibrium temperature $\sqrt{X}$. 
In vector notation, the explicit expression 
of the equilibrium Floquet level population is
\begin{eqnarray}
\boldsymbol{Q}(X)=
\frac{1}{\mathds{G}^{(0)}_{1,0}(X)+\mathds{G}^{(0)}_{0,1}(X)}
\begin{bmatrix}
\mathds{G}^{(0)}_{0,1}(X)
\\
\mathds{G}^{(0)}_{1,0}(X)
\end{bmatrix}.
\label{master:Q}
\end{eqnarray}
The function $F^{(0)}$ has  the interpretation of the leading order
contribution to the expansion in powers of $\varepsilon$ of the probability
density of the squared temperature $X$: 
\begin{eqnarray}
F(X,\tau)=\sum_{r=0,1} \bar{P}_{r}(X,\tau)=\sum_{n=0}^{\infty}\varepsilon^n\,F^{(n)}(X, \tau).
\label{master:marg}
\end{eqnarray}
In appendix~\ref{sec:effectT} we show that within $O(\varepsilon^{2})$ accuracy, the probability density
$F$ evolves according to the Fokker--Planck equation
\begin{eqnarray}
\label{eq:final}
\hspace{-0.4cm}\partial_{\tau} F(X,\tau)+\partial_{X}J(X)\, F(X, \tau)=
\partial^2_X\frac{S(X) {F}(X,\tau)}{2\,N}
\end{eqnarray}
with the drift
\begin{eqnarray}
J(X)=\frac{\Sigma\,V}{\gamma} \,(T_p^5 -X^{5/2})+\jmath^{(1)}(X)+\frac{\jmath^{(2)}(X)}{N}
\label{master:drift}
\end{eqnarray}
and positive definite diffusion coefficient
\begin{eqnarray}
S(X)=\frac{(\sqrt{10\Sigma V k_p}T_p^3 )^2}{\gamma^2}+\Delta^{(1)}(X)+\Delta^{(2)}(X)
\label{master:dc}
\end{eqnarray}
The $\jmath^{(i)}$, $\Delta^{(i)}$, $i=1,2$ terms embody the average effect of the 
fluctuating qubit-calorimeter energy flux close to equilibrium. Specifically, 
upon defining
\begin{eqnarray}
\boldsymbol{Z}=\begin{bmatrix}
1\\1
\end{bmatrix}
\label{master:zm}
\end{eqnarray}
we find that
\begin{subequations}
\label{master:j}
\begin{eqnarray}
\label{}
\label{master:j1}
&&\jmath^{(1)}(X)=-\langle\boldsymbol{Z}|\mathds{G}^{(1)}(X)\boldsymbol{Q}(X)\rangle
\\
&&
\jmath^{(2)}(X)=\frac{1}{\lambda(X)}\langle \boldsymbol{Z} |\mathds{G}^{(1)}(X)\mathds{J} \boldsymbol{Z} \rangle 
\langle\mathds{J}\boldsymbol{Q}(X)|\mathcal{L}_{X}^{(1)}\boldsymbol{Q}(X)\rangle
\nonumber\\&&\hspace{0.2cm}
+\frac{1}{\lambda(X)}\langle \boldsymbol{Z} |\mathds{G}^{(1)}(X) \mathds{J}\boldsymbol{Z} \rangle 
\left\langle\boldsymbol{Q}^{\perp}\big{|}\partial_{X}\big{(}\mathds{G}^{(1)}(X)\boldsymbol{Q}(X)\big{)}\right\rangle
\nonumber\\&&\hspace{0.2cm}
-\partial_{X}\left(\frac{\langle \boldsymbol{Z} |\mathds{G}^{(1)} \boldsymbol{V} \rangle 
\langle\boldsymbol{Q}^{\perp}|\mathds{G}^{(1)}(X)\boldsymbol{Q}(X)\rangle}{\lambda(X)}\right)
\label{master:j2}
\end{eqnarray}
\end{subequations}
where
\begin{eqnarray}
\lambda(X)=-\Big{(}\mathds{G}_{1,0}(X)+\mathds{G}_{0,1}(X)\Big{)}.
\label{master:lambda}
\end{eqnarray}
$\mathsf{J}$ is the $2\times 2$ symplectic matrix 
proportional to the $\sigma_{y}$ Pauli matrix
\begin{eqnarray}
\mathds{J}=-i\sigma_{y}=\begin{bmatrix}
0 & -1
\\
1 & 0
\end{bmatrix}
\label{master:symplectic}
\end{eqnarray}
and we use the $\mathbb{C}^{2}$ scalar product notation e.g.
\begin{eqnarray}
\langle\boldsymbol{Z}|\mathds{G}^{(1)}(X)\boldsymbol{Q}(X)\rangle\equiv\sum_{r,s=0}^{1}\mathds{G}^{(1)}_{r,s}(X)\,Q_{s}(X)
\nonumber
\end{eqnarray}
Similarly, we find
\begin{subequations}
\label{master:dc}
\begin{eqnarray}
\label{master:dc1}
&& \Delta^{(1)}(X)=\langle\boldsymbol{Z}|\mathds{G}^{(2)}(X)\boldsymbol{Q}(X)\rangle
\\
\label{master:dc2}
&&\Delta^{(2)}(X)
=2\,\frac{\langle \boldsymbol{Z} |\mathds{G}^{(1)}(X)\mathds{J} \boldsymbol{Z} \rangle 
\langle\mathds{J}\boldsymbol{Q}|\mathds{G}^{(1)}\boldsymbol{Q}(X)\rangle}{\lambda(X)}
\end{eqnarray}
\end{subequations}
In Appendix \ref{sec:effectT} we prove that the contributions (\ref{master:dc}) to the 
diffusion coefficient are indeed positive definite.  

The drift and diffusion coefficients (\ref{master:j}), (\ref{master:dc})
depend upon the detailed form of the potential driving the qubit.
At arbitrarily low temperatures and if the matrix elements (\ref{eq:jumpop2}) 
restrict the number of permitted transitions to $\mathcal{N}\sim O(1)$, we can nevertheless extricate some 
general properties of the diffusion process (\ref{eq:final}). Under these hypotheses, 
we expect that
\begin{eqnarray}
\jmath^{(1)}(X)+\jmath^{(2)}(X)=g^{2}O(\hbar \,\omega_{L})+O(N^{-1}).
\nonumber
\end{eqnarray}
Consequently the temperature probability distribution tends to a stationary value 
peaked around the temperature value at which the drift (\ref{master:drift}) vanishes
  \begin{eqnarray}
T_S^{5}\approx T_p^5+\frac{g^{2}}{\Sigma V}O(\hbar \omega_{L}^2).
\label{master:TS}
\end{eqnarray}
We assume that the terms on the right hand side are of the same order, as it occurs in the simulations in section \ref{sec:sim}.
The same line of reasoning suggests to capture the behavior of the bulk of the temperature 
distribution by means of the Ornstein--Uhlenbeck process obtained by setting
\begin{eqnarray}
J(X)\approx\frac{\mathrm{d}J }{\mathrm{d} X}(T_{S}^{2})\,(T_S^2-X).
\label{master:OUdrift}
\end{eqnarray}
and
\begin{eqnarray}
S(X)\approx S(T^{2}_{S}).
\label{master:OUdiff}
\end{eqnarray}
From the Ornstein--Uhlenbeck approximation we can immediately estimate the average steady state 
temperature as $T_{\star}\approx T_{S}$ and the relaxation time to equilibrium as
\begin{eqnarray}
\tau_{S}\approx \left(\frac{\mathrm{d}J }{\mathrm{d} X}(T_{S}^{2})\right)^{-1}
\label{master:rt}
\end{eqnarray}
Finally, neglecting completely thermal contributions to qubit-calorimeter energy exchanges
lead to infer the relation
\begin{eqnarray}
T_{S} \,\sim\,\left(T_p^5 + g^{2}\frac{O(\hbar\omega_{L}^2)}{\Sigma\,V}\right)^{1/5}
\label{master:scaling}
\end{eqnarray} 
between the peak of the equilibrium temperature distribution and the qubit-calorimeter coupling constant.
If we suppose that (\ref{master:j}) depends weakly on the temperature, (\ref{master:TS}) 
yields
\begin{eqnarray}
\tau_{S}&&\approx \left(\frac{5\,\Sigma\,V}{2\,\gamma}T_{S}^{3}\right)^{-1}\nonumber\\
&&\sim
\left(\frac{5\,\Sigma\,V}{2\,\gamma}\bigg(T_p^5 + g^{2}\frac{O(\hbar\omega_{L}^2)}{\Sigma\,V}\bigg)^{3/5} \right)^{-1}
\label{master:rtg}
\end{eqnarray}

\section{Simulations}
\label{sec:sim}

In order to obtain quantitative predictions, we consider in (\ref{model:qubit}) 
the driving potential 
\begin{equation}
V_d(t)=\hbar \omega_{q}\,(e^{i\omega_L t}\sigma_{+} +e^{-i\omega_L t}\sigma_{-}).
\end{equation}
The advantage of this choice \cite{BlBuGrSiSmWa91,GeKoSk95} is that we can derive analytic expressions 
for the Floquet states $\phi_{i,n}$ and the quasi energies $\epsilon_{i}$, $i=0,1$. The matrix elements (\ref{eq:jumpop2}) 
permit transitions corresponding to only six Lindblad operators \cite{BrPe97,BrPe02}
\begin{subequations}
\label{sim:Lindblad}
\begin{eqnarray}
\label{sim:Lindblad01}
&&\hspace{-0.6cm}A_{0,1}=\frac{\sin\theta}{2}\big{(}|\phi_{1,0}(0)
\rangle\langle\phi_{1,0}(0)|-|\phi_{0,0}(0)\rangle\langle\phi_{0,0}(0)|\big{)}
\\\label{sim:Lindblad11}
&&\hspace{-0.6cm}A_{1,1}=\sin^2\frac{\theta}{2}|\phi_{1,0}(0)\rangle\langle\phi_{0,0}(0)|
\\\label{sim:Lindblad-11}
&&\hspace{-0.6cm}A_{-1,1}=\cos^2\frac{\theta}{2}|\phi_{1,0}(0)\rangle\langle\phi_{0,0}(0)|
\end{eqnarray}
\end{subequations}
and their adjoint conjugates. In (\ref{sim:Lindblad}) we set
\begin{equation}
\cos\theta=\frac{\omega_{q}-\omega_{L}}{\nu}
\end{equation} 
with
\begin{equation}
\nu=\frac{|\epsilon_{0}-\epsilon_{1}|}{\hbar}=\sqrt{(\omega_q-\omega_L)^2+4\,\kappa^2\,\omega_{q}^{2}} 
\label{sim:gap}
\end{equation}   
The  operators (\ref{sim:Lindblad11}), (\ref{sim:Lindblad-11}) 
have the  same effect  on the  qubit but  describe respectively
the transfer of  $\hbar  \omega_L$ and  $-\hbar  \omega_L$ amounts of  energy
from the drive to the calorimeter through the qubit.
Inspection of (\ref{sim:gap}) also evinces that at resonance
\begin{eqnarray}
\omega_{q}=\omega_{L}
\nonumber
\end{eqnarray}
the condition securing the validity of the rotating wave approximation takes 
the particularly simple form \cite{BrPe02}
\begin{eqnarray}
\kappa\gg g^{2}.
\nonumber
\end{eqnarray}
Hence, the use of the Floquet representation of the qubit dynamics is well 
justified when the qubit is strongly coupled to the drive.

We integrate numerically the qubit-calorimeter dynamics for parameter 
values as in  \cite{SoAvPe13,HeNiPe08}. We take the level  spacing of
the  qubit  $\hbar\omega_q=k_{\mathrm{B}}\,\times 1 \mathrm{K}$,  the  volume of  the  calorimeter
$V=10^{-21}$   m\textsuperscript{3} ,   $\Sigma=2\times  10^{-9}$
WK\textsuperscript{-5}m\textsuperscript{-3} and the phonon temperature $T_p=0.1$ K.    Further     we    take
$\gamma=1500k_{B}/ (1$K)  and the drive coupling constant
$\kappa=0.05$.

At  the  beginning  of  the  simulations  the  driven  qubit  and  the
calorimeter are in thermal equilibrium  with the phonon
bath. The qubit  is in a thermal state at  temperature $T_p$. From the
thermal distribution we draw the initial Floquet state for the qubit.

We use the following algorithm for the numeric integration of the dynamics.   
We discretize time into steps of size $\diff t=(100 \omega_{q})^{-1}$. and update
the qubit state and temperature from  time $t$ to $t+\diff t$ in three
steps: (1) we compute the jump rates for the Poisson processes for
the qubit state $\psi$ and the temperature of the calorimeter $T_e$ at
time $t$,  (2) we let a random  number generator determine whether
the  qubit  makes a  jump  or  not, (3)  we update the  qubit  state $\psi$  and
temperature  $T_e$  using equations  \eqref{eq:sse}  and
\eqref{master:xi}. We repeat steps (1)-(3) for the duration of the qubit 
driving horizon.

We  study  the  temperature behavior  of the qubit-calorimeter system
 in two different regimes. We first look at a short time regime
of $10\times2\pi/\omega_q$.  In this regime  the qubit only  makes few  
jumps. Secondly,  we  look  at  the long  term  temperature
behavior.  After  waiting  sufficient time  the  temperature  process
converges towards a steady state.

\begin{figure}
\centering
\includegraphics[scale=0.4]{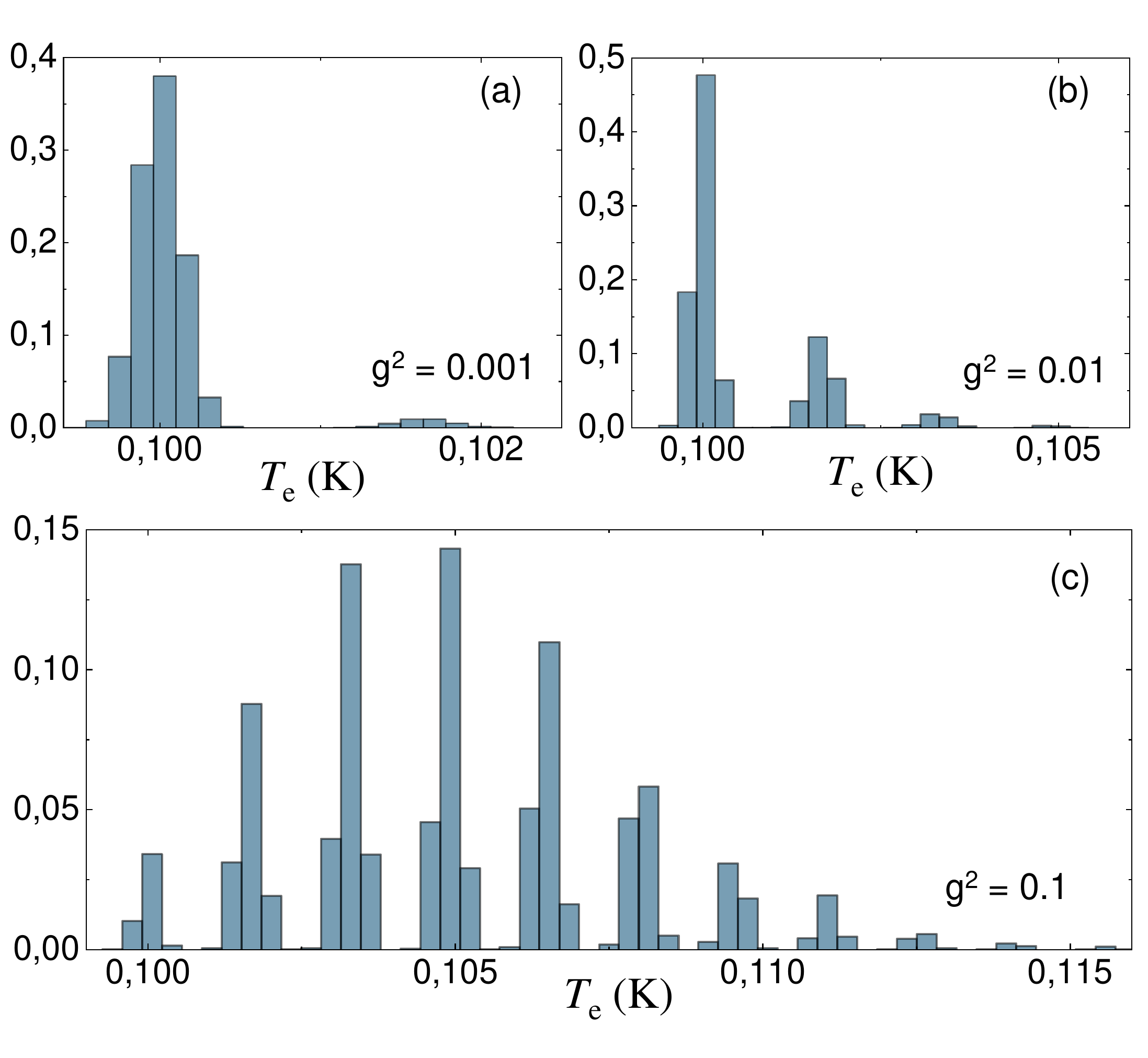}
\caption{Temperature distributions after 10 periods of resonant driving for different values of the qubit-calorimeter coupling $g$.
The distributions are obtained from histograms over $2\times 10^5$ realizations. 
The parameters used for the numerics are: $\hbar\omega_q=k_{\mathrm{B}}\times 1\mathrm{K}$,  $V=10^{-21}$ m\textsuperscript{3}, $\Sigma=2\times 10^{-9}$ 
WK\textsuperscript{-5}m\textsuperscript{-3}, $\gamma=1500k_B/(1$K), driving coupling constant $\kappa=0.05$  and the phonon temperature $T_p=0.1$ K.}
\label{fig:distr}
\end{figure}

Figure \ref{fig:distr} shows distribution  of the temperature after 10
periods of resonant driving. The temperature distributions are sharply
peaked  around values reachable via quantum jumps from  the  initial
temperature $T_p$. On  this time scale the dynamics is dominated by quantum jumps. 
Figure \ref{fig:distr}a shows how for low
coupling  the temperature  only makes few jumps.  As the  coupling
increases  more  jumps  occur.  The distribution  shifts  and  becomes
broader, see Figure \ref{fig:distr}c.

\begin{figure}
\centering
\includegraphics[scale=0.6]{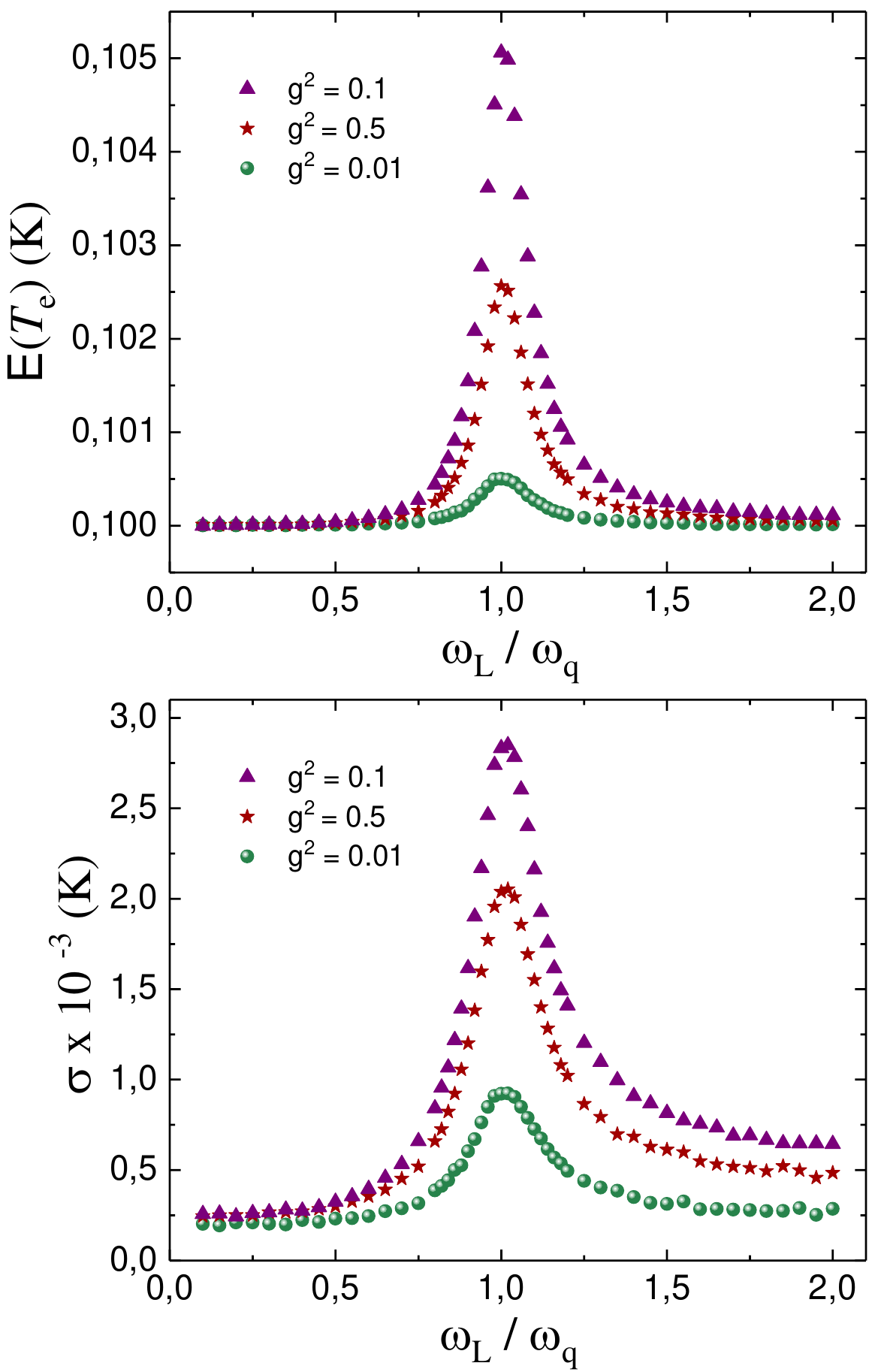}
\caption{The first moment (top) and the standard deviation (bottom) for the distribution of the temperature after $10\frac{2\pi}{\omega_q}$ duration of driving. Both the mean and standard deviation peak for resonant driving and higher coupling. The parameters used for the simulations are in the caption of Figure \ref{fig:distr}.}
\label{fig:short}
\end{figure}

Figure  \ref{fig:short}  shows the  first  and  second moment  of  the
distribution  of the  temperature  distributions like  those shown  in
Figure \ref{fig:distr} for different  driving frequencies. As expected, the average
temperature peaks around resonant frequency and is higher for stronger
coupling between the qubit and calorimeter.

\begin{figure}
\centering
\includegraphics[scale=0.3]{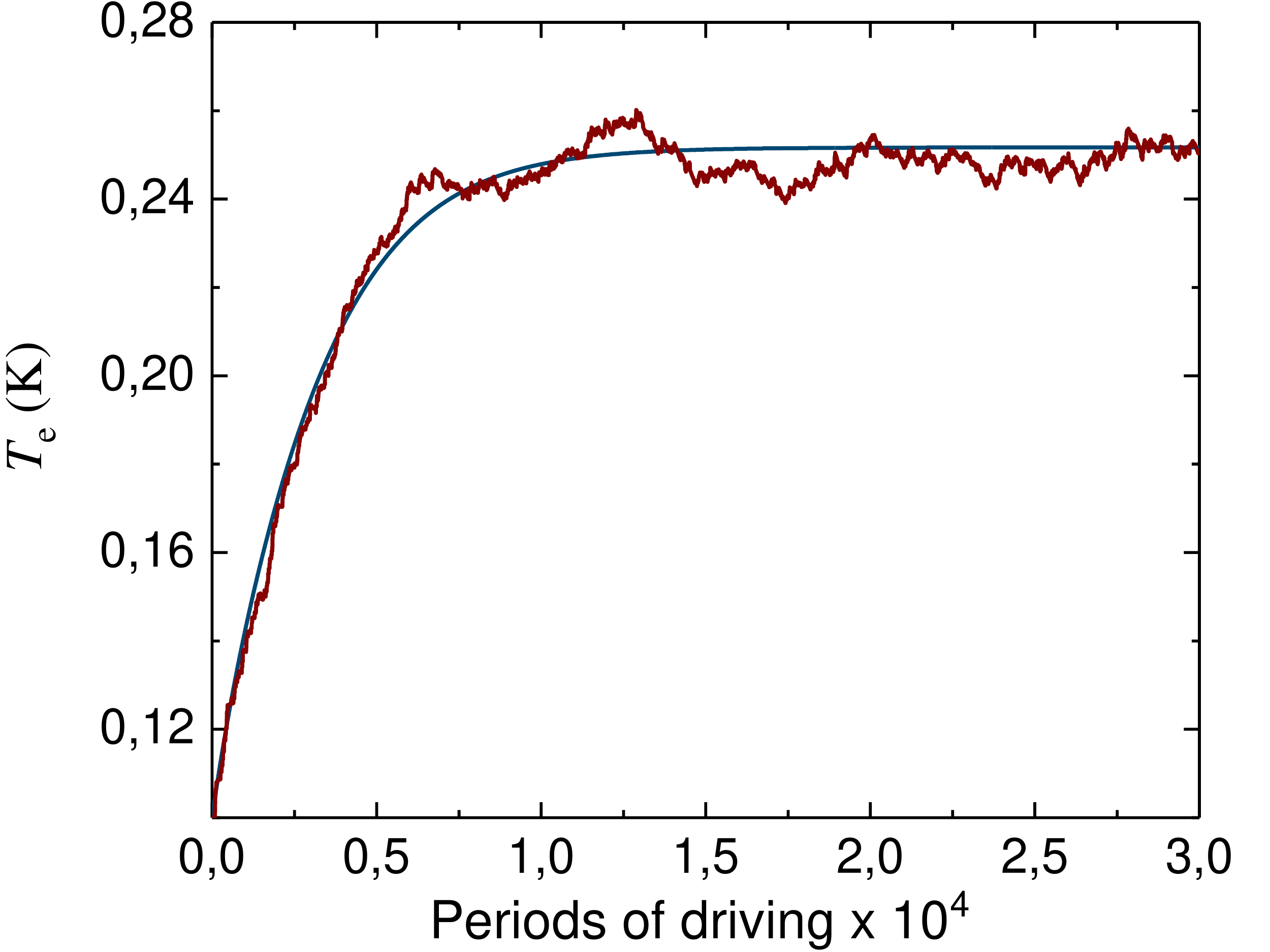}
\caption{The long time behaviour of the temperature, the noisy (red) line is a single realisation of the qubit-temperature process given by equations \eqref{eq:sse} and \eqref{eq:temptoE} for $g^2=1/100$. The smooth (blue) line is the evolution of the average temperature by the effective temperature process \eqref{eq:final}. The parameter values are the same as in 
Figure \ref{fig:distr}.}
\label{fig:longre}
\end{figure}

\begin{figure}
\centering
\includegraphics[scale=0.3]{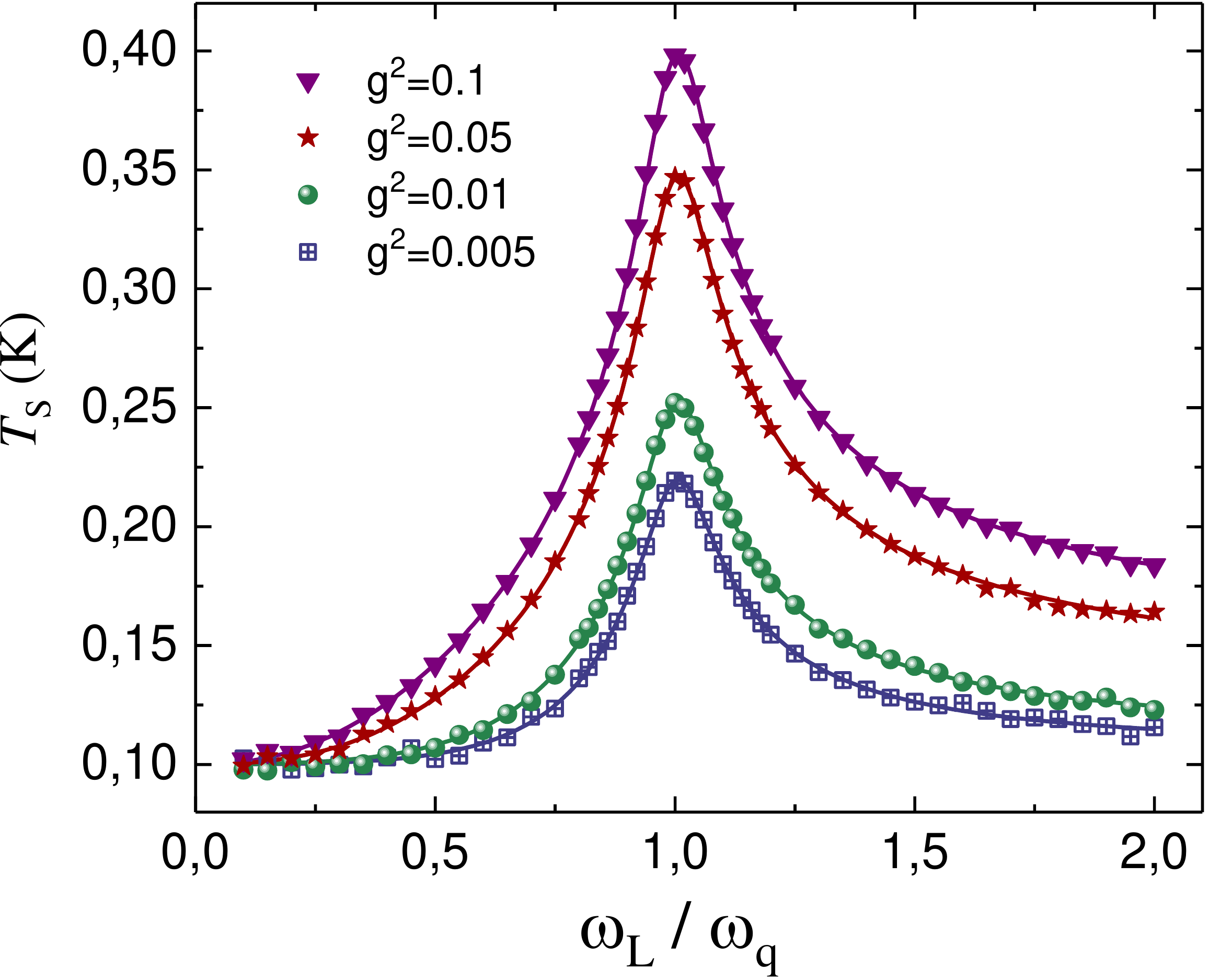}
\caption{Mean value of the temperature in the steady state. The data come from a single 
realization after it reached the steady state as shown in Figure \ref{fig:longre}. 
The full lines are the estimate of the stationary temperature obtained from 
the solution of \eqref{sim:drift}. The parameter values are the same as in 
Figure \ref{fig:distr}.}
\label{fig:long}
\end{figure}

On  timescales  of  the  order $10^4$  periods  the  qubit-temperature
process exhibits convergence towards a steady state. Figure~\ref{fig:longre} illustrates this
phenomenon. The (red) noisy line is a realisation of the qubit-temperature process. The smooth (blue) line is the evolution of the average temperature obtained from the analytic approximation, i.e. the evolution by the drift term of equation \eqref{eq:final}.
Figure  \ref{fig:long} shows the average value  of the temperature 
process in the steady state versus the driving frequency, which we use as an estimate for $T_S$.  
The full line  is an estimate of the same quantity as obtained by 
imposing the vanishing of the drift (\ref{master:drift}) and thus
solving numerically the transcendental equation
\begin{eqnarray}
J(X_{S})=0.
\label{sim:drift}
\end{eqnarray}
We notice that for $T_{e}=0$ the solution of this equation takes the form
\begin{eqnarray}
&&T_{S}^{5}=T_p^5+\frac{g^{2}}{\Sigma\,V}\bigg{(}\hbar \,\omega^2_L \sin^2(\theta) 
\nonumber\\
&&
+ \frac{\hbar(\omega_L+\nu)^{3}\sin^4(\theta/2)+\hbar(\omega_L-\nu)^{3}\cos^4(\theta/2)}
{(\omega_L+\nu)\sin^4(\theta/2)+(\omega_L-\nu)\cos^4(\theta/2)} 
\bigg{)}
\label{eq:monTs}
\end{eqnarray}
The  long  time   behavior  of  the  temperature  is  most
interesting  around  the  resonant  frequency. For  the  rest  of  our
numerical analysis we focus on resonant driving.

\begin{figure}
\centering
\centering
\includegraphics[scale=0.3]{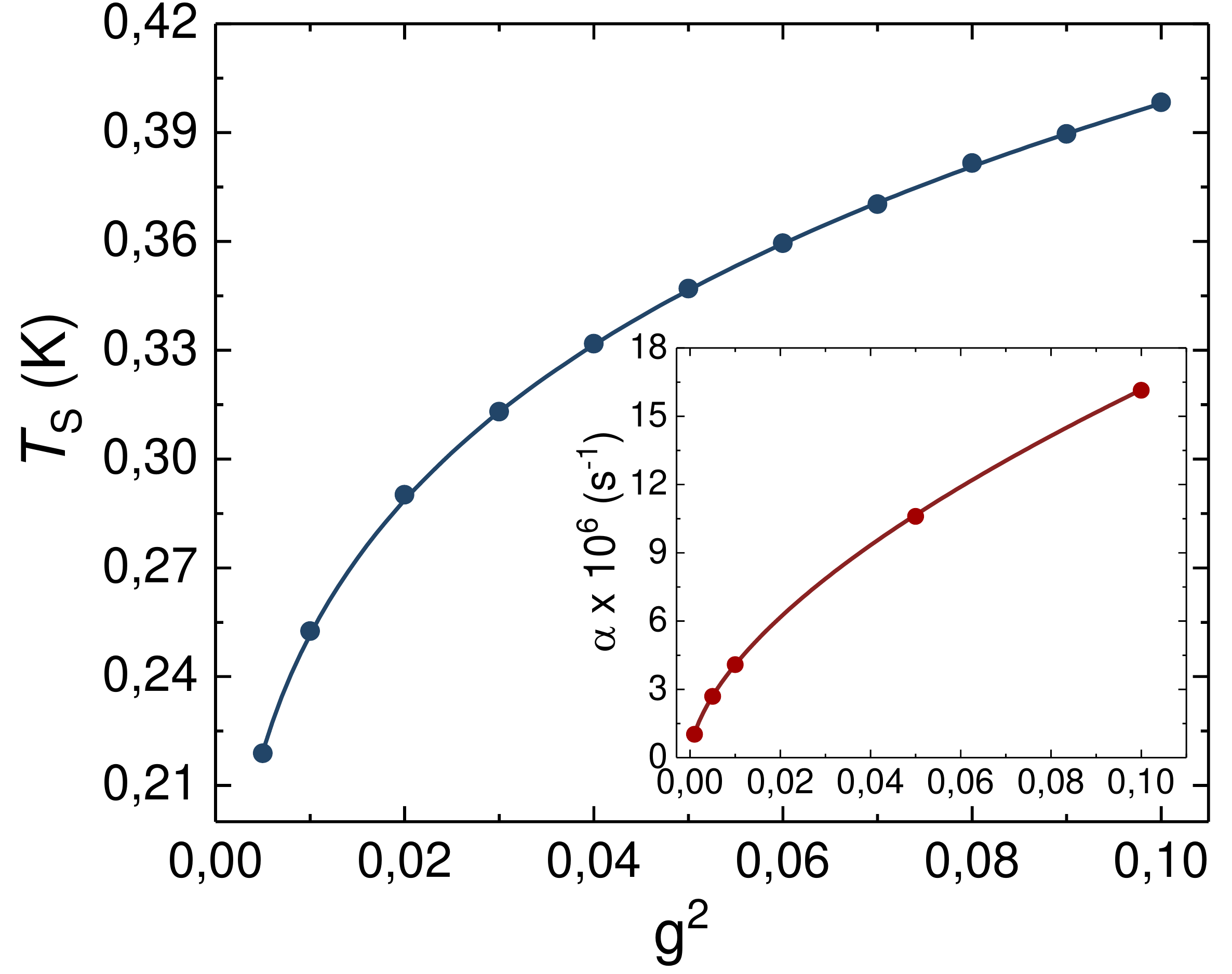}
\caption{The mean value of the temperature in the steady state for different values of $g^2$ at resonant driving. 
The (blue) dots give an estimate for the stationary temperature obtained from the simulations. 
The full (blue) line is the solution obtained from the analytic approximation,  the $g$ 
law from equation \eqref{eq:monTs}. (Inserted plot) The inverse of the relaxation time $\alpha=\tau_{S}^{-1}$. 
The (red) dots are obtained from the average of $10^3$ realisations of the qubit-calorimeter process by fitting to the temperature curve. The full (red) line is the $g$ law from equations \eqref{master:rtg} and \eqref{eq:monTs}.
The parameters used for the simulations are in the caption of Figure \ref{fig:distr}.}
\label{fig:statTa}
\end{figure}

In  Figure \ref{fig:statTa}  we compare  the value for $T_S$ from equation (\ref{eq:monTs}) 
(full  line)  with the average steady state temperature obtained from direct numerical 
simulations (dots).  We find good agreement with  the $g$ dependence   predicted  
by (\ref{master:scaling}).  
Furthermore, we compare the relaxation time prediction of the Ornstein--Uhlenbeck approximation
with the numeric observation. The inserted plot in Figure \ref{fig:statTa} shows $\alpha=\tau_S^{-1}$, it demonstrates that the data are consistent with the
$g$ dependence predicted by (\ref{master:rtg}) and (\ref{eq:monTs}).

\begin{figure}
\centering
\includegraphics[scale=0.3]{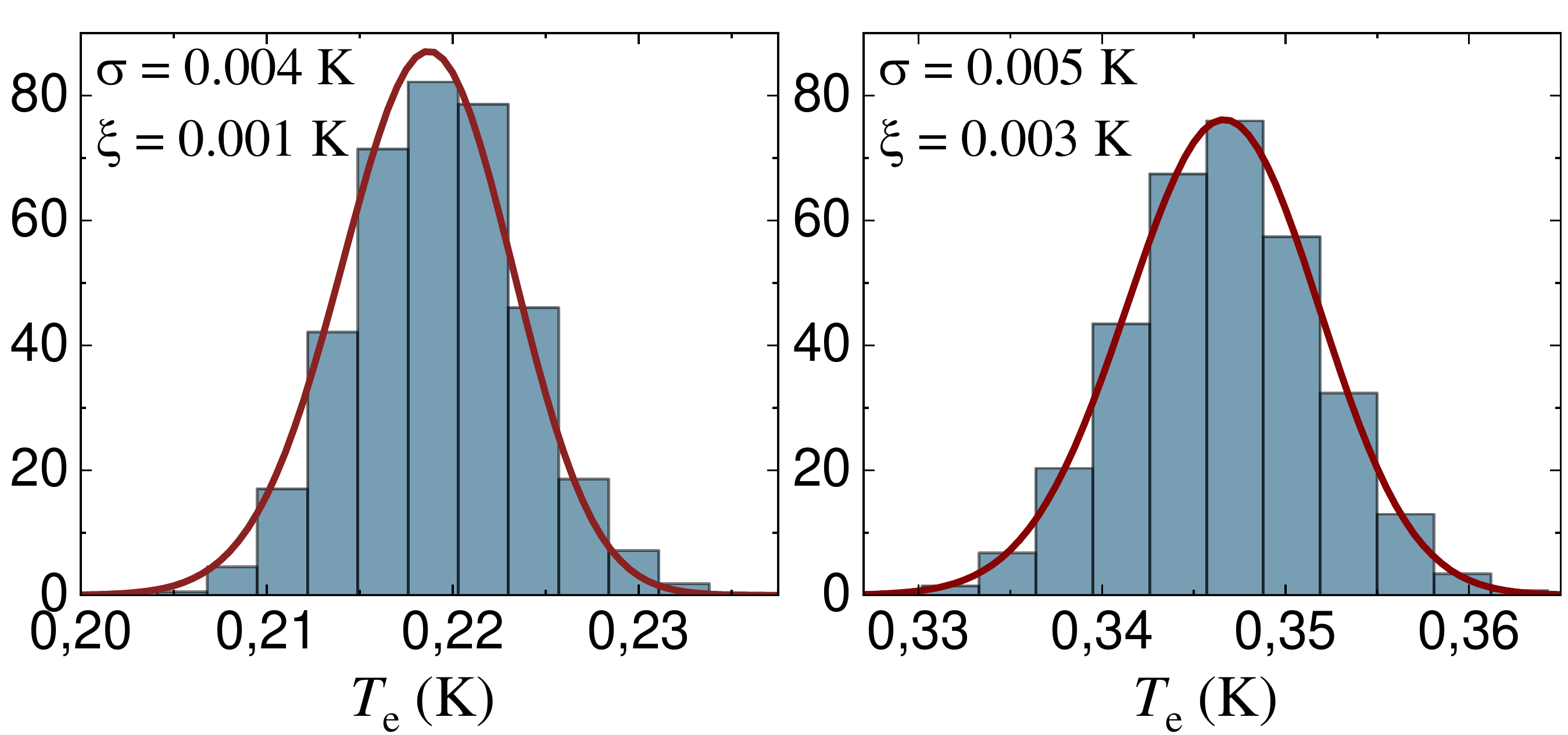}\\
\setlength{\unitlength}{0.1cm} 
\begin{picture}(0,0)
\put(-15.7,37.6){$g^2=0.005$}
\put(27,37.6){$g^2=0.05$}
\end{picture}
\caption{Probability density functions for the qubit-temperature steady 
state for different values of the coupling $g^2$.
The (blue) histogram is generated from the numerical simulations. The full (red) line is the 
solution of equation \eqref{eq:final}. The values for the variance $\sigma=(\mathsf{E}[T_e(t)-\mathsf{E}(T_e(t))]^2)^{1/2}$ and the skewness $\xi=(\mathsf{E}[T_e(t)-\mathsf{E}(T_e(t))]^3)/\sigma^3$ obtained from the numerics are given in the figures. The parameters used for the simulations are in the 
caption of Figure \ref{fig:distr}. }
\label{fig:distrr}
\end{figure}
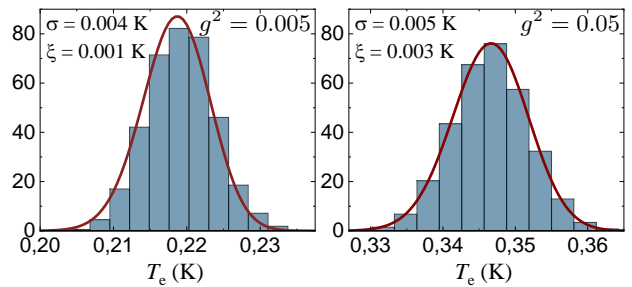

In Figure \ref{fig:distrr} we plot the stationary value of the temperature for 
different values of the qubit-electron coupling $g$. We construct the histograms 
by sampling a single realization of the qubit-temperature process after convergence to the 
steady state. The full (red) line is the stationary solution of the Fokker--Planck equation 
\eqref{eq:final}.  In Figure \ref{fig:distrr} we 
also report the values of the standard  deviation and  skewness as obtained from the numerics.  
In the stationary state  the average value $T_{\star}$ of the temperature is 
close  to the temperature  $T_{S}$ specified by the solution of (\ref{sim:drift}). 
The  square root  of the variance of the temperature process ranges 
from $0.004$ K to $0.005$ K.

Finally,  Figure~\ref{fig:power} shows  a  log-log plot  of the  power
spectrum of the temperature process. We obtain the data by following the evolution
of a single realization of the temperature process after it has reached  
the steady state.  The spectrum exhibits a decay consistent with a fit equal to $-2$ of the slope. 
This is  in agreement with the Ornstein--Uhlenbeck approximation (\ref{master:OUdrift}), (\ref{master:OUdiff}) 
of the drift and diffusion coefficients in the Fokker--Planck equation
(\ref{eq:final}).  We find in such a case the expression of the power spectrum
\begin{eqnarray}
\mathcal{S}(\omega)=\frac{S^{2}(T_{S}^{2})\,\tau_{S}^{2}}{1+\omega^{2}\,\tau_{S}^{2}}
\nonumber
\end{eqnarray}
where $\tau_{S}$ is the relaxation time of the process.

\begin{figure}
\centering
\includegraphics[scale=0.35]{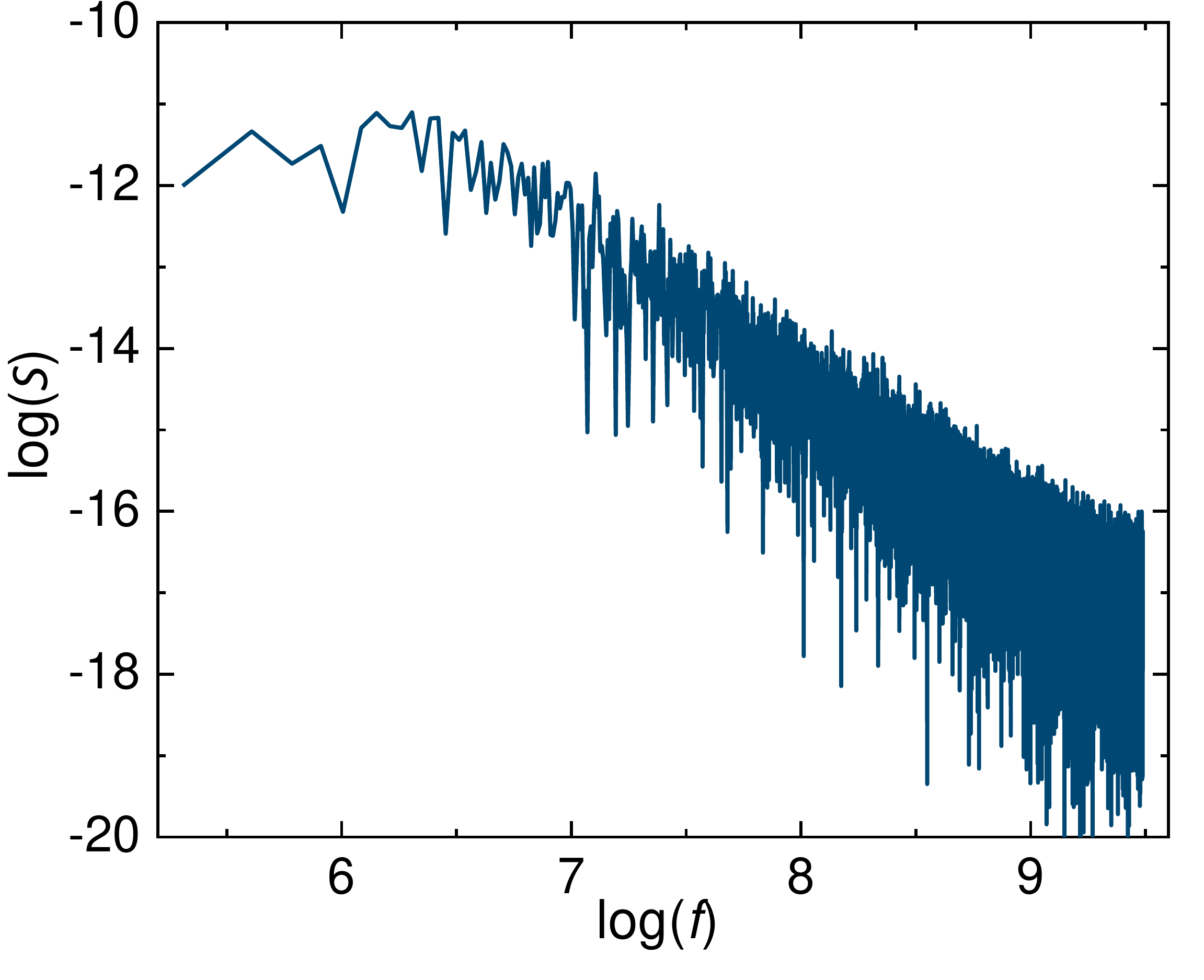}
\label{fig:power}
\caption{Power spectrum, for $g^2=\frac{10}{100}$. The spectrum decays with a $\omega^{-2}$ 
asymptotic law consistent with the Ornstein--Uhlenbeck approximation (\ref{master:OUdrift}), (\ref{master:OUdiff})}
\label{fig:power}
\end{figure}

\section{Conclusion and Outlook} 

In   summary,  we   present  a   theoretical  model   of  calorimetric
measurements  in   an integrated   quantum  circuit   consisting  of  a
superconducting qubit and a normal  metal absorber element.  The joint
evolution of  the population  of the  qubit state  and the
calorimeter temperature is governed  by the Chapman--Kolmogorov master
equation  (\ref{master:master}).    Standard  methods   of  asymptotic
analysis reduce this equation  to an effective Fokker--Planck equation
for  the  probability  distribution  of  the  calorimeter  temperature
alone. In  the asymptotic regime,  we are able to  make experimentally
testable predictions about the dependence of statistical indicators of
temperature fluctuations upon the qubit-calorimeter coupling constant.

The  engineering   of  quantum   integrated  circuits   of  increasing
tunability  is in  a phase  of rapid  development \cite{CoJeBrCaHu17,MaFuUeNa17,PaTaAlNiMo17}.  In particular, very
recently \cite{RoKaSeCh18}  has shown  the realizability of  a quantum
heat valve to  observe tunable heat transport  between mesoscopic heat
reservoirs at different temperatures. The laboratory implementation is
a   resonator-qubit-resonator  assembly   in   which   the  qubit   is
capacitively embedded  between two superconducting  transmission lines
each  terminated  by  a  normal  metal  resistor  elements  acting  as
mesoscopic heat reservoirs at different temperatures. The study of the
heat flow  in the  presence of  resonator elements  thus appears  as a
natural direction towards which extend to the ideas of the present work.

\section{Acknowledgments}

We  warmly  thank Bayan  Karimi  for  discussions  and help  with  the
graphics.  We  are also  gratefully acknowledge discussions  with Lara
Ul\v{c}akar, Antti Kupiainen and Dmitry Golubev.  The work of B. D. is
supported by DOMAST.  B.~D. and P.M-G. also acknowledge support by the
Centre  of Excellence  in  Analysis  and Dynamics  of  the Academy  of
Finland. The work of J. P. P. is funded through Academy of Finland grant 312057 and from the European Union's Horizon 2020 research and innovation programme under the European Research Council (ERC) programme (grant agreement 742559).

\appendix

\section{Time scales in the model}
\label{ap:Floquet}

Let $H(t)=H(t+\mathsf{T}_{\mathrm{p}})$ be a $\mathsf{T}_{\mathrm{p}}$-periodic self-adjoint matrix 
acting on $\mathbb{C}^{d}$. Floquet theory see e.g. \cite{Shi65,Zel67,Sam73,BlBuGrSiSmWa91,BrPe97,GrHa98,Hol15,Don18} 
links up solutions of the initial value problem
\begin{eqnarray}
\begin{cases}
\Big{(}H(t)-\imath\,\hbar\,\partial_t\Big{)}\psi(t)=0
\nonumber\\
\psi(0)=\bar{\psi}
\end{cases}
\label{Floquet:ode}
\end{eqnarray}
with the spectral problem
\begin{equation}
\label{Floquet:spectral}
\begin{cases}
\Big{(}H(t)-\imath\,\hbar\,\partial_t\Big{)}\phi_{r,n}(t)=\epsilon_{r,n}\phi_{r,n}(t)
\nonumber\\
\phi_{r,n}(t+\mathsf{T}_{\mathrm{p}})=\phi_{r,n}(t).
\end{cases}
\end{equation}  
in the Hilbert space $\mathcal{H}=\mathbb{C}^{d}\times L^2[0,\mathsf{T}_{\mathrm{p}}]$.
Namely, if we denote by $\mathsf{F}$ the fundamental solution of 
(\ref{Floquet:ode})
\begin{eqnarray}
\psi(t)=\mathsf{F}(t,0)\bar{\psi}\,,\hspace{1.0cm}\forall\,\psi\,\in\,\mathbb{C}^{d}
\label{Floquet:flow}
\end{eqnarray}
and by $\left\{\boldsymbol{e}_{r}\right\}_{r=1}^{d}$ the orthonormal basis (Floquet's states) in $\mathbb{C}^{d}$ 
diagonalizing the monodromy matrix  
\begin{eqnarray}
\mathsf{F}(\mathsf{T}_{\mathrm{p}},0)\boldsymbol{e}_{r}=
e^{-\frac{\imath\,\epsilon_{r}\,\mathsf{T}_{\mathrm{p}}}{\hbar}}\,\boldsymbol{e}_{r}
\label{monodromy:Fs}
\end{eqnarray}
then, for $r=1,\dots,d$ and $n\in\mathbb{Z}$ the identities
\begin{subequations}
\label{Floquet:spectrum}
\begin{eqnarray}
\label{}
&&\phi_{r,n}(t)=e^{\frac{\imath}{\hbar}\,\left(\epsilon_{r}+\frac{2\,\pi\,n}{\mathsf{T}_{\mathrm{p}}}\right)t}\,F_{t,0}\boldsymbol{e}_{r}
\label{Floquet:eigenvectors}
\\
&&\epsilon_{r,n}=\epsilon_{r}+\frac{2\,\pi\,n}{\mathsf{T}_{\mathrm{p}}}
\label{Floquet:eigenvalues}
\end{eqnarray}
\end{subequations}
solve  the spectral problem (\ref{Floquet:spectral}).
The eigenvalues (\ref{Floquet:eigenvalues}) are the quasi-energies, 
see eq. (\ref{quT:quasi}) in the main text. The eigenvectors
(\ref{Floquet:eigenvectors}) form a complete basis of $\mathcal{H}$. 
Setting the quantum number $n$ to zero conventionally specifies the first Brillouin 
zone. Note also that
\begin{eqnarray}
\phi_{r,n}(0)=\boldsymbol{e}_{r}
\label{Floquet:ev}
\end{eqnarray}
for all $r,n$.

An immediate consequence of the completeness of the $\phi_{r,n}$'s is 
that any solution of (\ref{Floquet:ode}) admits the expression
\begin{eqnarray}
\psi(t)=\sum_{r=1}^{d}\sum_{n\in\mathbb{Z}}
\phi_{r,n}(t)\,e^{-\frac{\imath\,\epsilon_{r,n}\,t}{\hbar}}\langle\langle \phi_{r,n}\,|\,\bar{\psi}\rangle\rangle
\label{Floquet:expansion}
\end{eqnarray}
In (\ref{Floquet:expansion}) $\langle\langle\dots\rangle\rangle$ is the widely adopted physics notation
for scalar product over $\mathcal{H}$ i.e. for any $f\,,g\,,\in \mathcal{H}$ 
\begin{eqnarray}
\langle\langle \phi_{r,n}\,|\,\bar{\psi}\rangle\rangle\equiv\langle f\,,g\rangle_{\mathcal{H}}
=\int_{0}^{\mathsf{T}_{\mathrm{p}}}\frac{\mathrm{d}t}{\mathsf{T}_{\mathrm{p}}}
\langle f(t)\,|\,g(t)\rangle
\nonumber
\end{eqnarray}
whereas 
\begin{eqnarray}
\langle f(t)\,|\,g(t)\rangle=\langle f(t)\,,g(t)\rangle_{\mathbb{C}^{d}}
\nonumber
\end{eqnarray}
is the usual Dirac's notation for the scalar product over $\mathbb{C}^{d}$.
Finally, the insertion in (\ref{Floquet:flow}) of the completeness relation in $\mathbb{C}^{d}$ 
in terms of the Floquet basis $\left\{\boldsymbol{e}_{r}\right\}_{r=1}^{d}$
combined with the definition (\ref{Floquet:eigenvectors}) of 
eigenstates of the spectral problem in the first Brillouin zone yields 
the identity
\begin{eqnarray}
\psi(t)
=\sum_{r=1}^{d}
\phi_{r,0}(t)\,e^{\frac{\imath\,\epsilon_{r,0}t}{\hbar}}\,\langle \boldsymbol{e}_{r}|\bar{\psi}\rangle
\nonumber
\end{eqnarray}
This is the so-called Floquet's representation of solutions of (\ref{Floquet:ode}).
As the coefficients $\langle \boldsymbol{e}_{r}|\bar{\psi}\rangle$ 
do not depend upon time, their absolute square value admits the interpretation of
population probability of the Floquet state $r$. 
See \cite{GrHa98,Hol15,Don18} for details.

\section{Qubit-Electron interaction}
\label{ap:qe}

Let us consider the closed qubit-calorimeter dynamics. The Dirac's picture
Hamiltonian is
\begin{eqnarray}
\tilde{H}_{q e}=\mathsf{F}^{\dagger}(t,0)\,e^{\frac{\imath \,H_{e}\,t}{\hbar}}
\,H_{q e}\,e^{-\frac{\imath\,H_{e}\,t}{\hbar}}\mathsf{F}(t,0)
\label{qe:interaction}
\end{eqnarray}
with $\mathsf{F}$ the flow (\ref{Floquet:flow}).
The Hamiltonian is the sum of tensor products of operators independently 
acting on the Hilbert space of the qubit and of the electrons.
The operator acting on the qubit Hilbert space always admits the representation
\begin{eqnarray}
\lefteqn{
\mathsf{F}^\dagger(t,0)(\sigma_++\sigma_-)\mathsf{F}(t,0)
}
\nonumber\\&&
=\sum_{r,s=0}^{1} e^{\imath \frac{\epsilon_{r,0} -\epsilon_{s,0}}{\hbar}t}
|\phi_{r,0}(0)\rangle \tilde{D}_{r,s}(t)  \langle \phi_{s,0}(0)|
\nonumber
\end{eqnarray}
where 
\begin{eqnarray}
\tilde{D}_{r,s}(t)=\langle\phi_{r,0}(t)|\sigma_{+}+\sigma_{-}|\phi_{s,0}(t)\rangle
\nonumber
\end{eqnarray}
The completeness for any $t$ in $\mathbb{C}^{2}$ of the Floquet basis immediately implies
\begin{eqnarray}
\tilde{D}_{0,0}(t)=-\tilde{D}_{1,1}(t)
\nonumber
\end{eqnarray}
Furthermore, $\tilde{D}_{r,s}(t)$ is a $2\,\pi/\omega_{L}$ periodic function the Fourier series whereof 
is amenable to the form
\begin{eqnarray}
\tilde{D}_{r,s}(t)=\sum_{n\in\mathbb{Z}}e^{\imath\,\omega_{L}\,n\,t}D_{r,s,n}
\label{qe:expansion}
\end{eqnarray}
with $D_{r,n,s}$ defined by (\ref{eq:jumpop2}). The advantage of the Floquet 
representation is to couch the time dependence of the Dirac picture 
Hamiltonian into the form of a sum over purely oscillating exponentials 
as in the case of bipartite isolated systems.

In the weak coupling scaling limit, at leading order we consider 
transition occurring for non-vanishing matrix elements of (\ref{qe:interaction})
satisfying the resonance condition
\begin{eqnarray}
\eta_{k}-\eta_{l}=\epsilon_{r,0} -\epsilon_{s,0}+n\,\hbar\,\omega_{L}
\nonumber
\end{eqnarray}
where $\eta_{k}$, $\eta_{l}$ are energy levels of the free electron Hamiltonian. These considerations 
\cite{BrPe02} fix the form of the Lindblad operators (\ref{eq:jumpop1}).

Finally, to explain the Bose--Einstein distribution appearing in
(\ref{eq:gam}), we observe that the emission of $\hbar \omega$ energy 
from the qubit to the calorimeter occurs with rate
\begin{eqnarray} 
\lefteqn{\hspace{-0.4cm}
\mathcal{R}(\omega)\propto
}
\nonumber\\&&\hspace{-0.4cm}
\frac{g^{2}}{N^{2}}\sum_{i\,j}f_{T_{e}}(\eta_{i})\left(1-f_{T_{e}}(\eta_{j})\right)
\frac{\sin\big{(}\frac{\eta_{i}-\eta_{j}-\hbar\omega}{\hbar}t\big{)}}
{\eta_{i}-\eta_{j}-\hbar\omega}
\end{eqnarray}
where $t$ is the duration of the interaction,  $\eta_{i}$ denotes the $i$-th electron energy level and
\begin{equation}
f_{T_{e}}(\eta)=\frac{1}{e^{(\eta-\mu)/(k_{B}\,T_{e})}+1}.
\label{qe:FD}
\end{equation}
is the Fermi--Dirac distribution at temperature $T_{e}$.
In the large $N$ limit, we approximate the double sum over the electron energy levels
with a double integral. The integrand is then amenable to further simplifications.
The weak coupling scaling limit yields
\begin{eqnarray}
\frac{\sin\big{(}\frac{\eta_{i}-\eta_{j}-\hbar\omega}{\hbar}t\big{)}}
{\eta_{i}-\eta_{j}-\hbar\omega}\overset{t \uparrow \infty}{\to}\,\hbar\,\pi\,
\delta\left(\eta_{i}-\eta_{j}-\hbar\omega\right)
\nonumber
\end{eqnarray}
Moreover, the low temperature limit permits to set
the energy density of states to a constant value in the region where the integrand is
sensibly different from zero \cite{InNa92}. Finally we can extend the range of integration
to the full real axis. The upshot is
\begin{eqnarray}
\mathcal{R}(\omega)\propto g^{2}\int_{\mathbb{R}}\mathrm{d}E\,f_{T_{e}}(E)\bigg{(}1-f_{T_{e}}(E-\hbar\,\omega)\bigg{)} 
\nonumber
\end{eqnarray}
We avail us of the identity
\begin{eqnarray}
\lefteqn{
f_{T_{e}}(E)\bigg{(}1-f_{T_{e}}(E-\hbar\,\omega)\bigg{)}
}
\nonumber\\&&
=\frac{e^{\hbar\,\omega/(K_{B}T_{e})}}{e^{\hbar\,\omega/(K_{B}T_{e})}-1}\bigg{(}f_{T_{e}}(E-\hbar\,\omega)-f_{T_{e}}(E)\bigg{)}
\label{qe:identity}
\end{eqnarray}
to couch the integral into the form
\begin{align}
\mathcal{R}(\omega)\approx \frac{ g^2\,e^{\hbar\omega/(k_{B}T_{e})}}{e^{\hbar\omega/(k_{B}T_{e})}-1} 
\, \int_{\mathbb{R}} \mathrm{d}E\, (f(E-\hbar\,\omega)-f(E))
\end{align}
and upon noticing that
\begin{eqnarray}
\frac{\mathrm{d} \mathcal{R}}{\mathrm{d} \omega}(\omega)
\approx-
\frac{ g^2\,e^{\hbar\omega/(k_{B}T_{e})}}{e^{\hbar\omega/(k_{B}T_{e})}-1} 
\, \int_{\mathbb{R}} \mathrm{d}E\, \frac{1}{\hbar}\frac{\mathrm{d} }{\mathrm{d} E}f(E-\hbar\,\omega)
\nonumber
\end{eqnarray}
we finally get into
\begin{eqnarray}
\mathcal{R}(\omega)\approx \frac{ g^2\,\hbar\,\omega\,e^{\hbar\omega/(k_{B}T_{e})}}{e^{\hbar\omega/(k_{B}T_{e})}-1} 
\nonumber
\end{eqnarray}

\section{Electron-Phonon interaction}
\label{sec:appPh}

For reader convenience, we summarize here the calculation of the first
two moments  of the energy  flux between  the phonon and  the electron
reservoirs. We perform the calculation under the following hypotheses 
\cite{WeUrCl94}
\begin{enumerate}
\item[i] The electron gas 
\begin{eqnarray}
H_{e}=\sum_{k}\eta_{k}\,c_{k}^{\dagger}\,c_{k}
\nonumber
\end{eqnarray}
is initially at equilibrium at a uniform temperature $T_{e}\ll T_{F}$ 
with $T_{F}$ the Fermi temperature. The energy of an electron 
having wave-number $\boldsymbol{k}$ is
\begin{eqnarray}
\eta_{k}=\frac{\hbar k^{2}}{2\,m}\,,\hspace{1.0cm}k=\|\boldsymbol{k}\|
\nonumber
\end{eqnarray}
\item[ii] 
The phonon gas 
\begin{eqnarray}
H_{p}=\sum_{k}\hbar\,\omega_{k}\,b_{k}^{\dagger}\,b_{k}
\nonumber
\end{eqnarray}
is initially at equilibrium with an uniform temperature $T_p\,\ll\,T_{D}$ 
with $T_{D}$ the Debye temperature \cite{AsMe76}. 
In this temperature limit, phonons obey a linear dispersion relation
\begin{eqnarray}
\omega_k=v_s\,k
\nonumber
\end{eqnarray}
$v_{s}$ the speed of sound and $k=\|\boldsymbol{k}\|$ for $\boldsymbol{k}$ the phonon wavelength.
\item[iii] The  interaction between the phonons  and the electrons
in the material is given by
\begin{equation}
H_{e p}=\kappa\sum_{\boldsymbol{k},\boldsymbol{q}} \omega_{q}^{1/2}
\Big{(}\,c^{\dagger}_{\boldsymbol{k}}\,c_{\boldsymbol{k}-\boldsymbol{q}}\,b_{\boldsymbol{q}}
+c^{\dagger}_{\boldsymbol{k}}\,c_{\boldsymbol{k}-\boldsymbol{q}}\,b^{\dagger}_{\boldsymbol{q}}
\Big{)}
\label{appPh:Hpe}
\end{equation}
The sum in (\ref{appPh:Hpe}) ranges over energies sufficiently close to the Fermi surface.
\item[iv] Scattering processes with out-coming phonons with wave numbers 
in a different Brillouin zone than incoming ones, are negligible 
(no ``umklapp'' \cite{AsMe76}).
\item[v] The dimensions of the metal are much longer than
the average phonon wavelength. This means that sums over wave numbers can be replaced
by integrals over approximately \emph{constant} density of states $D$ for phonons and $N$ for electrons.
\end{enumerate}
Following \cite{PeKa18} we evaluate the average heat current
in terms of the current operator $\mathfrak{J}$ defined by
\begin{eqnarray}
J=\frac{\mathrm{d} }{\mathrm{d} t}\operatorname{Tr}\left(\frac{H_{e}-H_{p}}{2}\,\boldsymbol{\rho}_{t}\right)\equiv
\operatorname{Tr} (\mathfrak{J}\,\boldsymbol{\rho}_{t}
)\label{appPh:currents}
\end{eqnarray}
Here $\boldsymbol{\rho}_{t}$ is the state operator of the phonon-electron system 
in Schr\"odinger's picture. The Liouville--von Neumann equation
yields
\begin{eqnarray}
\mathfrak{J}=-\frac{\imath\,\kappa}{2\,\hbar}\sum_{k,q}\omega_{q}^{1/2}\,\Omega_{k,q}\,
(a^{\dagger}_{\boldsymbol{k}}\, a_{\boldsymbol{k}-\boldsymbol{q}}\, c_{\boldsymbol{q}}
- a^{\dagger}_{\boldsymbol{k}-\boldsymbol{q}}\, a_{\boldsymbol{k}}\, c^{\dagger}_{\boldsymbol{q}} )
\nonumber
\end{eqnarray}
with $\Omega_{k,q}=\omega_{q}+\eta_{k}-\eta_{\|\boldsymbol{k}-\boldsymbol{q}\|}$. 
Turning to Dirac's picture and writing $\tilde{\mathfrak{J}}$ for heat current in said picture, within leading order accuracy in the weak 
coupling limit \cite{BrPe02} the  average heat current
\begin{eqnarray}
J=\imath\operatorname{Tr}\int_{0}^{t}\mathrm{d}s[\tilde{H}_{p e}(s),\tilde{\mathfrak{J}}(t)]\boldsymbol{\rho}_{0}+\mathrm{h.o.t.}
\nonumber
\end{eqnarray}
is amenable \cite{KaLiTa57,WeUrCl94} to the difference $J=J_{a}-J_{e}$ 
of two terms physically corresponding to the absorption and the emission 
of one phonon by the electron gas. Under the aforementioned hypotheses \textbf{i}-\textbf{v}, 
the absorption term is \cite{WeUrCl94}
\begin{eqnarray}
J_{a}=C\int\diff^3 q\,n_{T_{p}}(\omega_{q})\,\Big{(}n_{T_{e}}(\omega_{q})+1\Big{)}\,\hbar^{2}\omega_{q}^{2}\,I(\omega_{q})
\label{appPh:absorption}
\end{eqnarray}
whilst emission is
\begin{eqnarray}
J_{e}=C\int\diff^3 q\,\Big{(}n_{T_{p}}(\omega_{q})+1\Big{)}\,n_{T_{e}}(\omega_{q})\,\hbar^{2}\omega_{q}^{2}\,I(\omega_{q})
\label{appPh:emission}
\end{eqnarray}
with
\begin{eqnarray}
\lefteqn{\hspace{-0.3cm}
I(\omega_{q})=
}
\nonumber\\&&\hspace{-0.3cm}
\int\diff^3 k \Big{(}f_{T_{e}}(\eta_{k-q})-f_{T_{e}}(\eta_{k})\big{)}
\,\delta(\eta_{k}-\eta_{\|\boldsymbol{k}-\boldsymbol{q}\|}-\hbar\omega_q)
\label{appPh:fermion}
\end{eqnarray}
In writing (\ref{appPh:absorption}), (\ref{appPh:emission}) we defined
$C=2\,\pi\,D\,N\,\kappa^{2}\hbar^{-1}$
and we took advantage of 
the explicit form of the Fermi--Dirac (\ref{qe:FD}) and Bose--Einstein distributions
\begin{equation}
n_{T_{p}}(\omega_q)=\frac{1}{e^{\hbar\omega_q/(k_{B}T_{p})}-1}
\nonumber
\end{equation}
and of the identity (\ref{qe:identity}).
We also exploited the fact that the Dirac delta in (\ref{appPh:fermion}) fixes 
 the difference $\eta_{\|\boldsymbol{k}-\boldsymbol{q}\|}-\eta_{k}=\omega_{q}$ to a $\boldsymbol{k}$ independent value.
The integral (\ref{appPh:fermion}) is most conveniently evaluated in polar coordinates
\begin{eqnarray}
\lefteqn{
I(\omega_{q})= 
}
\nonumber\\&&
2\,\pi\,\int_{0}^{\infty}\hspace{-0.2cm}\diff k\,k^2\int_{-1}^1\diff z\,\iota(k,z;\omega_{q})
\,\delta\left(\hbar\,\omega_q+\frac{\omega_{q}\, k\, z}{m\,v_{s}}-\eta_{q}\right)
\nonumber
\end{eqnarray}
where $z$ is the angle between $\boldsymbol{k}$ and $\boldsymbol{q}$, and
\begin{eqnarray}
\lefteqn{\iota(k,z;\omega_{q})=}
\nonumber\\&&
f_{T_{e}}\left(\eta_{k}+\eta_{q}-\frac{\omega_{q}\, k\, z}{m\,v_{s}}\right)-f_{T_{e}}(\eta_k)
\nonumber
\end{eqnarray}
Upon evaluating the integral over $z$ we find 
\begin{eqnarray}
\lefteqn{
I(\omega_{q})=
}
\nonumber\\&&
\frac{2\,\pi\, m^2\,v_{s}}{\hbar^4 \omega_{q}}
\int_{E_{min}}^{+\infty}\diff E 
\,\Big{(}f_{T_{e}}(E-\hbar\omega_q)-f_{T_{e}}(E)\Big{)}
\nonumber
\end{eqnarray}
having set $E=\frac{\hbar^2 |k|^2}{2m}$ and
\begin{eqnarray}
E_{min}=\frac{\hbar^2}{2m}\left(\frac{\omega_{q}}{2\,v_{s}}+\frac{v_s m}{\hbar}\right)^2
\nonumber
\end{eqnarray}
The remaining integrand is peaked around $\mu$. 
Under our working hypotheses (see \cite{KaLiTa57,WeUrCl94}), the chemical potential
satisfies $\mu\gg\frac{\hbar^2}{2m}(\frac{q}{2}+\frac{v_s m}{\hbar})^2$ allowing us to write
\begin{eqnarray}
\lefteqn{
I(\omega_{q})\approx
}
\nonumber\\&&
\frac{2\,\pi\, m^2\,v_{s}}{\hbar^4 \omega_{q}}
\int_{-\infty}^{+\infty}\diff E \,\Big{(}f_{T_{e}}(E-\hbar\omega_q)-f_{T_{e}}(E)\Big{)}
\nonumber
\end{eqnarray} 
whence 
\begin{eqnarray}
I(\omega_{q})\approx\frac{2\,\pi\, m^2\,v_{s}}{\hbar^3}
\nonumber
\end{eqnarray}
We thus get into
\begin{eqnarray}
\lefteqn{
J=J_{a}-J_{e}=
}
\nonumber\\&&
\frac{2\,\pi m^2\,C\,v_{s}}{\hbar^3}
\int\diff^3 q \,\Big{(}n_{T_{p}}(\omega_{q})-n_{T_{e}}(\omega_{q})\Big{)}\,\omega_q^{2}
\end{eqnarray}
The remaining integral is the proportional to the difference between two 
averages with respect to the Bose--Einstein distribution. It can be evaluated 
by standard techniques see e.g. \cite{AsMe76}. The final result is
\begin{equation}
J=\Sigma\, V\,(T_p^5-T_e^5)
\label{appPh:mean}
\end{equation}
where $V$ is the volume of the metal and \cite{PeKa18}
\begin{equation}
\Sigma =\frac{12\,\kappa^{2}\,\zeta(5)\,m\,k_{\mathrm{B}}^{5}}{\pi\,k_{F}\,v_{s}^{2}\,\hbar^{6}}
\label{appPh:sigma}
\end{equation}
with $\zeta$ the Riemann zeta functions an $k_{F}$ the Fermi momentum. 
The definition of $\Sigma$ hinges upon setting $D=V/(2\pi)^{3}$ for the phonon density 
of states

The evaluation of current correlation function 
\begin{eqnarray}
\mathcal{C}_{t}=\operatorname{Tr}U^{\dagger}_{t}\mathfrak{J} U_{t}\mathfrak{J}\boldsymbol{\rho}_{0}
\nonumber
\end{eqnarray}
with 
\begin{eqnarray}
U_{t}=\exp\left(-\imath\frac{H_{e}+H_{p}+H_{p e}}{\hbar} t\right)
\nonumber
\end{eqnarray}
proceeds along the same lines as above. We refer to \cite{PeKa18} for details.
Within leading accuracy and at $T_e=T_p$ we get into
\begin{equation}
\int_{-\infty}^{\infty}\mathrm{d}t\, \mathcal{C}_{t}=10\,\Sigma\, V\, k_B\,T_p^6.
\end{equation}
We use this result to weight Brownian fluctuations in the temperature process.

\section{Master equation}
\label{sec:apMaster}

In this Appendix we derive the master equation (\ref{master:master}). We start by writing 
the probability (\ref{master:prob}) in the form
\begin{equation}
\label{apMaster:prob}
P_{r}(X,t)=\mathsf{E}\Big{(}|\langle \boldsymbol{e}_{r}|\psi(t)\rangle|^2\delta(\xi(t)-X)\Big{)}
\end{equation}
where $\mathsf{E}(.)$ is the average and $\boldsymbol{e}_{r}=\phi_{r,0}(0)$. We find the master equation by evaluating
\begin{equation}
\label{eq:dp}
\diff P_{r}(X,t)=\mathsf{E}\diff\Big{(}|\langle \boldsymbol{e}_{r}|\psi(t)\rangle|^2\delta(\xi(t)-X)\Big{)}
\end{equation}
Let us call $f(\psi_t,\psi^*_t,\xi)=|\langle\boldsymbol{e}_{r}|\psi\rangle|^2\delta(\xi-X)$. 
The differential of $f$ is
\begin{eqnarray}
\lefteqn{
\diff f(\psi,\psi^*,\xi)\equiv
}
\nonumber\\&&
f(\psi+\diff\psi,\psi^*+\diff\psi^*,\xi+\diff \xi)- f(\psi,\psi^*,\xi)=
\nonumber\\&&
\sum_{\substack{p=1\\p=k_1+k_2+k_3}}^{\infty}\hspace{-0.5cm}
\frac{(\diff \xi)^{k_1}(\diff \psi^*)^{k_2}(\diff \psi)^{k_3}}{k_1!k_2!k_3!}
\partial_{\xi}^{k_1}\partial_{\psi^*}^{k_2}\partial_{\psi}^{k_3}f(\psi,\psi^*,\xi)
\nonumber
\end{eqnarray}
We then use \eqref{eq:sse} and \eqref{master:xi} to express
the differentials $\mathrm{d}\psi$, $\mathrm{d}\psi^{*}$ and $\mathrm{d}\xi$, in terms of the time differential $\mathrm{d}t$ 
and the increments $\mathrm{d}w$ and $\mathrm{d}\nu$ of Wiener and Poisson processes. 
The rules of stochastic calculus, see e.g. \cite{Jac10}, impose
$\diff w^{2}(t)=t$, $\diff w(t)\diff \nu_{r,n}(t)=0$ and $\diff \nu_{r,n}(t) \diff \nu_{r^{\prime},n^{\prime}}(t)
=\delta_{r,r^{\prime}}\delta_{n,n^{\prime}}\diff \nu_{r,n}(t)$. 
We thus get into the It\^o--Poisson stochastic differential
\begin{eqnarray}
\lefteqn{
\diff f(\psi,\psi^*,\xi)=\mathcal{L}_{\xi}^{\dagger}f(\psi,\psi^*,\xi)\diff t
}
\nonumber\\&&
+\frac{\sqrt{10\Sigma V k_B}T^3_p}{\gamma}\partial_{\xi} f(\psi,\psi^*,\xi)\diff w(t)
\nonumber\\&&
-\frac{\imath}{\hbar}\Big{(} \,G(\psi)\partial_{\psi}-G^{*}(\psi)\partial_{\psi^*}\Big{)}f(\psi,\psi^*,\xi)\diff t
\nonumber\\&&
+\mathrm{d}_{jump}f(\psi,\psi^*,\xi)
\label{apMaster:df}
\end{eqnarray}
$\mathcal{L}^{\dagger}$ is the $\mathbb{L}^{(2)}$ adjoint of \eqref{master:FP} with respect to 
the Lebesgue measure
\begin{eqnarray}
\mathcal{L}_{\xi}^{\dagger}f=\frac{\Sigma\,V}{N\,\gamma}(T^5_p-\xi^{5/2})\partial_{\xi}f
+\frac{10\Sigma V k_BT^6_p}{2\,N^{2}\,\gamma^2}\partial^2_{\xi} f
\nonumber
\end{eqnarray}
Furthermore we can couch
\begin{eqnarray}
\lefteqn{
-\frac{\imath}{\hbar}\Big{(}G(\psi)\partial_{\psi}-G^{*}(\psi)\partial_{\psi^*}
\Big{)}\,|\left\langle\boldsymbol{e}_{r}\right|\psi\rangle|^{2}\delta(\xi-X)
=}
\nonumber\\&&
\sum_{\substack{|s|\leq 1\\|n|\leq\mathcal{N}}}\Gamma(\omega_{s,n},\xi)
\|A_{s,n}\psi\|^2|\langle\boldsymbol{e}_{r}|\psi\rangle|^{2}\delta(\xi-X)
\nonumber\\&&
-\sum_{\substack{|s|\leq 1\\|n|\leq\mathcal{N}}}\Gamma(\omega_{s,n},\xi)\operatorname{Re}
\langle\boldsymbol{e}_{r}|A^{\dagger}_{s,n}A_{s,n}|\psi\rangle
\langle\psi|\boldsymbol{e}_{r}\rangle\delta(\xi-X)
\nonumber
\end{eqnarray}
into the form
\begin{eqnarray}
\lefteqn{
-\imath\,\hbar\Big{(}G(\psi)\partial_{\psi}-G^{*}(\psi)\partial_{\psi^*}
\Big{)}\|\left\langle\boldsymbol{e}_{r}\right|\psi\rangle\|^{2}\delta(\xi-X)
=}
\nonumber\\&&
\sum_{\substack{|s|\leq 1\\|n|\leq\mathcal{N}}}\Gamma(\omega_{s,n},\xi)\,\|A_{s,n}\psi\|^{2}
\langle\boldsymbol{e}_{r}|\psi\rangle|^2\delta(\xi-X)\diff t
\nonumber\\&&
-\sum_{\substack{r^{\prime}=0,1\\|n|\leq\mathcal{N}}}\mathcal{G}_{r^{\prime},r,n}(\xi)\,
\langle\boldsymbol{e}_{r}|\psi\rangle|^2\delta(\xi-X)\diff t
\label{apMaster:Pdrift}
\end{eqnarray}
having used (\ref{eq:jumpop1}), (\ref{eq:jumpop2}) to derive
\begin{eqnarray}
\lefteqn{
\sum_{|s|\leq 1}
\Gamma(\omega_{s,n},\xi)\operatorname{Re}(
\langle\boldsymbol{e}_{r}|A^{\dagger}_{s,n}A_{s,n}|\psi\rangle\langle\psi,\boldsymbol{e}_{r}\rangle)
}
\nonumber\\&&
=\Gamma(\omega_{0,n},\xi)|D_{1,1,n}|^{2}
\langle\boldsymbol{e}_{r}|\psi\rangle
\nonumber\\&&
+\Big{(}\delta_{r,0}\Gamma(\omega_{1,n},\xi)\,|D_{1,0,n}|^{2}
+\delta_{r,1}\Gamma(\omega_{-1,n},\xi)\,|D_{0,1,n}|^{2}
\Big{)}
\langle\boldsymbol{e}_{r}|\psi\rangle
\nonumber
\end{eqnarray}
and the definition (\ref{master:rate2}) for $\mathcal{G}_{r^{\prime},r,n}(\xi)$. 
The last term on the right hand side of (\ref{apMaster:df}) is purely due to jumps
\begin{eqnarray}
\lefteqn{
\mathrm{d}_{jump}f(\psi,\psi^*,\xi)=
-\sum_{\substack{s=-1\\|n|\leq\mathcal{N}}}^{1}\mathrm{d}\nu_{s,n}f(\psi,\psi^*,\xi)
}
\nonumber\\&&
+\sum_{\substack{|s|\leq 1\\|n|\leq\mathcal{N}}}\mathrm{d}\nu_{s,n}f\left(\frac{(A_{s,n}\psi)}{\|A_{s,n}\psi\|},
\frac{(A_{s,n}\psi)^*}{\|A_{s,n}\psi\|},\xi+\frac{\hbar\omega}{N\,\gamma}\right)
\nonumber
\end{eqnarray}
or more explicitly
\begin{eqnarray}
\lefteqn{
\mathrm{d}_{jump}f(\psi,\psi^{*},\xi)=
}
\nonumber\\&&
-\sum_{\substack{s=-1\\|n|\leq\mathcal{N}}}^{1}\mathrm{d}\nu_{s,n}\,
\langle\boldsymbol{e}_{r}|\psi\rangle|^2\delta(\xi-X)\diff t
\nonumber\\&&
+\sum_{\substack{s=-1\\|n|\leq\mathcal{N}}}^{1}\mathrm{d}\nu_{s,n}\,\frac{|\langle\boldsymbol{e}_r|A_{s,n}|\psi\rangle|^2}
{\|A_{s,n}\psi\|^{2}}
\delta\left(\xi+\frac{\hbar\omega_{s,n}}{N\,\gamma}-X\right)
\label{apMaster:pj}
\end{eqnarray}
Taking the expectation value of (\ref{apMaster:df}) brings about 
several simplifications. To start with, the term proportional to 
the increment of the Wiener vanishes owing to the It\^o prescription
\cite{Jac10} whereas the identity
\begin{eqnarray}
\mathsf{E}\Big{(}|\langle\boldsymbol{e}_{r}|\psi_t\rangle|^2\mathcal{L}_{\xi}\delta(\xi-X)\Big{)}
=\mathcal{L}_{X}P_{r}(X,t)
\nonumber
\end{eqnarray}
holds in consequence of the properties of the Dirac-$\delta$ distribution. 
By (\ref{eq:condav}), the expectation value of (\ref{apMaster:pj}) yields
\begin{eqnarray}
\lefteqn{
\mathsf{E}\Big{(}\mathrm{d}_{jump}f(\psi,\psi^{*},\xi)\Big{)}=
}
\nonumber\\&&
-\sum_{\substack{s=-1\\|n|\leq\mathcal{N}}}^{1}\mathsf{E}\Big{(}\Gamma(\omega_{s,n},\xi)\,\|A_{s,n}\psi\|^{2}
\langle\boldsymbol{e}_{r}|\psi\rangle|^2\delta(\xi-X)\Big{)}\diff t
\nonumber\\&&
+\sum_{\substack{r^{\prime}=0,1\\|n|\leq\mathcal{N}}}\mathsf{E}\left(\mathcal{G}_{r,r^{\prime},n}(\xi)\,
\langle\boldsymbol{e}_{r^{\prime}}|\psi\rangle|^2\delta\left(\xi+\frac{\hbar\,\omega_{r-r^{\prime},n}}{N\,\gamma}-X\right)\right)\diff t
\nonumber
\end{eqnarray}
having also used (\ref{eq:jumpop1}) to evaluate
\begin{eqnarray}
\lefteqn{
|\langle\boldsymbol{e}_r|A_{s,n}|\psi\rangle|^2=\delta_{s,0}|D_{1,1,n}|^{2}\,|\langle\boldsymbol{e}_{r}|\psi\rangle|^{2}
}
\nonumber\\&&
+(\delta_{s,1}\delta_{r,1}+\delta_{s,-1}\delta_{r,0})|D_{r,r-s,n}|^{2}|\langle\boldsymbol{e}_{r-s}|\psi\rangle|^{2}
\nonumber
\end{eqnarray} 
and the definition (\ref{master:rate2}) of the rates of the master equation. If we contrast 
this last result with (\ref{apMaster:Pdrift}) we notice that the first term on the
right hand side of both expression mutually cancel.
Gathering all non vanishing contributions, and recalling the definitions (\ref{master:rate1}), 
(\ref{master:rate2}) we obtain
\begin{eqnarray}
\lefteqn{
\frac{\mathrm{d} }{\mathrm{d} t}P_{r}(X,t)=\mathcal{L}_{X}P_{r}(X,t)
}
\nonumber\\&&
+\sum_{\substack{r^{\prime}=0,1\\|n|\leq\mathcal{N}}}\hspace{-0.2cm}\mathcal{G}_{r, r^{\prime}, n}\left(X-\frac{\hbar\,\omega_{r-r^{\prime},n}}{N\,\gamma}\right)
\,P_{r^{\prime}}\left(X-\frac{\hbar\,\omega_{r-r^{\prime},n}}{N\,\gamma},t\right)
\nonumber\\&&
-\sum_{\substack{r^{\prime}=0,1\\|n|\leq\mathcal{N}}}\mathcal{G}_{r^{\prime}, r, n}(X)
\,P_{r}(X,t)
\end{eqnarray}
which is \eqref{master:master}.

\section{Temperature process}
\label{sec:effectT}

We analyze here the perturbative solution of (\ref{master:multiscale})
up to order $O(\varepsilon^{2})$.

\paragraph{Order $\varepsilon^{0}$}

The lowest order satisfies
\begin{subequations}
\label{effectT:zero}
\begin{eqnarray}
\label{effectT:zero1}
&&\sum_{s=0}^{1}\bigg(\mathds{G}^{(0)}_{r,s}(X)Q_{s}(X)-\mathds{G}^{(0)}_{s, r}(X)Q_{r}(X)\bigg)
=0
\\
\label{effectT:zero2}
&&Q_{0}(X)+Q_{1}(X)=1
\end{eqnarray}
\end{subequations}
It is helpful to represent the condition (\ref{effectT:zero1}) in the matrix form
\begin{eqnarray}
\mathds{M}(X)\boldsymbol{Q}(X)=0
\nonumber
\end{eqnarray}
where $\mathds{M}$ is the two dimensional matrix
\begin{eqnarray}
\mathds{M}(X)
=
\begin{bmatrix}
-\mathds{G}^{(0)}_{1,0}(X) & \mathds{G}^{(0)}_{0,1}(X)
\\
\mathds{G}^{(0)}_{1,0}(X) & -\mathds{G}^{(0)}_{0,1}(X)
\end{bmatrix}
\label{effectT:M}
\end{eqnarray}
As required by probability conservation, columns of (\ref{effectT:M})
add up to zero. The solution of (\ref{effectT:zero1}) is the thermal state 
for the qubit at temperature $T=\sqrt{X}$
\begin{equation}
\label{effectT:thermalstate}
Q_{r}(X)=\frac{\mathds{G}^{(0)}_{r,1-r}(X)}{\mathds{G}^{(0)}_{1,0}(X)+\mathds{G}^{(0)}_{0,1}(X)}\,\hspace{0.5cm} r=0,1
\end{equation}
which in vector notation is (\ref{master:Q}).

\paragraph{Order $\varepsilon$}

The first order correction solves
\begin{align}
\label{effectT:ffirstorder}
&\sum_{s=0}^{1}\mathds{M}_{r s}(X)\bar{P}^{(1)}_{s}(X,t)=-\dot{F}^{(0)}(X,t)Q_{r}(X)
\nonumber\\
&+\sum_{s=0}^{1}\big{(}\mathcal{L}_{X}^{(1)}\delta_{r s}+\partial_{X}\mathds{G}^{(1)}_{r s}(X)\big{)}\,Q_{s}(X)\,F^{(0)}(X,t)
\end{align}
By Fredholm's alternative \cite{PaSt08}, linear non-homogeneous
equations of generated by an Hilbert's expansion are solvable if
the non-homogeneous term is orthogonal to the kernel of the adjoint
$\mathds{M}^{\dagger}$ of the leading order linear operator $\mathds{M}$ \cite{PaSt08}.

The spectral analysis of $\mathds{M}$ shows that the dual zero mode equation
\begin{eqnarray}
\mathds{M}^{\dagger}\boldsymbol{Z}=0
\nonumber
\end{eqnarray}
yields (\ref{master:zm}). We choose the normalization of $\boldsymbol{Z}$ 
such that (\ref{effectT:zero2}) can be re-written as the scalar product
\begin{eqnarray}
\langle\boldsymbol{Z}|\boldsymbol{Q}\rangle\equiv\sum_{r=0}^{1}Z_{r}Q_{r}(X)=1
\label{effectT:kernnorm}
\end{eqnarray}
The quantity $\lambda$ introduced in (\ref{master:lambda}) is the non vanishing eigenvalue of
$\mathds{M}$, $\mathds{M}^{\dagger}$. The corresponding left eigenvector is
\begin{subequations}
\label{effectT:ortho}
\begin{eqnarray}
\label{effectT:ortho1}
&&\mathds{M}^{\dagger}(X)\boldsymbol{Q}^{\perp}(X)=\lambda(X)\boldsymbol{Q}^{\perp}(X)
\\\label{effectT:ortho2}
&&\boldsymbol{Q}^{\perp}(X)=\frac{1}{\lambda(X)}
\begin{bmatrix}
\mathds{G}^{(0)}_{1,0}
\\
-\mathds{G}^{(0)}_{0,1}
\end{bmatrix}
=\mathds{J}\boldsymbol{Q}
\end{eqnarray}
\end{subequations}
with $\mathds{J}$ defined by (\ref{master:symplectic}), so that
\begin{eqnarray}
\sum_{r=0}^{1}Q^{\perp}_{r}(X)Q_{r}(X)=\langle\mathds{J}\boldsymbol{Q}|\boldsymbol{Q}\rangle=0
\nonumber
\end{eqnarray}
as $\mathds{J}$ is real antisymmetric. The right eigenvector is
\begin{subequations}
\label{}
\begin{eqnarray}
\label{}
&&\mathds{M}(X)\boldsymbol{V}(X)=\lambda(X)\boldsymbol{V}(X)
\\
&&\boldsymbol{V}(X)=
\begin{bmatrix}
-1
\\
1
\end{bmatrix}=\mathds{J}\,\boldsymbol{Z}
\end{eqnarray}
\end{subequations}
normalized so that
\begin{eqnarray}
\sum_{r=0}^{1}Q^{\perp}_{r}(X)V_{r}(X)=1
\nonumber
\end{eqnarray}
Finally we notice that for any $X$ we can write the completeness relation in $\mathbb{C}^{2}$
of left and right eigenvectors of $\mathds{M}$ as
\begin{eqnarray}
\mathds{1}=|\boldsymbol{Q}\rangle\langle\boldsymbol{Z}|+|\boldsymbol{V}\rangle\langle\boldsymbol{Q}^{\perp}|
\label{effectT:completeness}
\end{eqnarray}  
Projecting (\ref{effectT:ffirstorder}) onto the zero mode (\ref{master:zm}), 
yields the solvability condition
\begin{eqnarray}
\label{eq:F0}
\lefteqn{
\dot{F}^{(0)}(X,t)	
=
}
\nonumber\\&&
\mathcal{L}_{X}^{(1)}\,F^{(0)}(X,t)+\partial_X\jmath^{(1)}(X)\,F^{(0)}(X,t)
\end{eqnarray}
with $\jmath^{(1)}$ respectively defined by (\ref{master:marg}) and \eqref{master:j1}. 
This equation determines $F^{(0)}$. From the probabilistic point of view $F^{(0)}$
is within leading order approximation the probability density for the squared temperature
$X$. From the geometric slant, $F^{(0)}$ is, within the same accuracy, the coordinate 
in the $\boldsymbol{Q}$, $\boldsymbol{V}$ basis of the solution of (\ref{master:multiscale}):
\begin{eqnarray}
F^{(0)}=\langle \boldsymbol{Z}|\boldsymbol{P}^{(0)}\rangle
\nonumber
\end{eqnarray}
The projection of \eqref{effectT:ffirstorder} onto (\ref{effectT:ortho2}) 
yields
\begin{eqnarray}
\lambda\sum_{r=0}^{1}Q^{\perp}_{r}\bar{P}_{r}^{(1)}=
\sum_{r,s=0}^{1}Q^{\perp}_{r}(\mathcal{L}_{X}^{(1)}\delta_{r s}+\partial_{X}\mathds{G}^{(1)}_{r s})Q_{s}
\,F^{(0)}
\nonumber
\end{eqnarray}
This equation yields the component along $\boldsymbol{V}$ of 
\begin{eqnarray}
\boldsymbol{\bar{P}}^{(1)}=F^{(1)}\,\boldsymbol{Q}+F_{V}^{(1)}\boldsymbol{V}
\nonumber
\end{eqnarray}
where
\begin{subequations}
\label{}
\begin{eqnarray}
\label{}
F^{(1)}&=&\langle\boldsymbol{Z}|\boldsymbol{P}^{(1)}\rangle
=\bar{P}^{(1)}_{0}(X,t)+\bar{P}^{(1)}_{1}(X,t)
\nonumber
\\
F_{V}^{(1)}&=&\frac{\langle\boldsymbol{Q}^{\perp}|\boldsymbol{P}^{(1)}\rangle}{\lambda}
\nonumber\\
&=&\frac{1}{\lambda}\left\langle\boldsymbol{Q}^{\perp}\big{|}(\mathcal{L}_{X}^{(1)}\boldsymbol{Q})\,F^{(0)}
+\partial_{X}(\mathds{G}^{(1)}\boldsymbol{Q}\,F^{(0)})\right\rangle
\nonumber
\end{eqnarray}
\end{subequations}

\paragraph{Order $\varepsilon^2$}

The second order equation is 
\begin{eqnarray}
\label{eq:secondorder}
\lefteqn{
\sum_{s=0}^{1}\mathds{M}_{r s}\bar{P}^{(2)}_{s}=-\partial_{t}\bar{P}^{(1)}_{r}+ \mathcal{L}_{X}^{(1)}\bar{P}^{(1)}_{r}
}
\nonumber\\&&
+\sum_{s=0}^{1}\partial_{X}(\mathds{G}^{(1)}_{r s} P^{(1)}_{s})
+\mathcal{L}_{X}^{(2)}Q_{r}F^{(0)}+\sum_{s=0}^{1} \partial_{X}(\mathds{G}^{(2)}_{r s} Q_{s}F^{(0)})
\nonumber
\end{eqnarray}
The solvability condition is
\begin{eqnarray}
\lefteqn{
\partial_{t}F^{(1)}=\mathcal{L}_{X}^{(1)}F^{(1)}+\mathcal{L}_{X}^{(2)}F^{(0)}
}
\nonumber\\&&
+\sum_{r,s=0}^{1}\left( \partial_{X}\mathds{G}^{(1)}_{r s} P^{(1)}_{s}
+\frac{1}{2}\partial_{X}^{2}\mathds{G}^{(2)}_{r s}Q_{s}\,F^{(0)}\right)
\nonumber
\end{eqnarray}
or equivalently in the scalar product notation
\begin{eqnarray}
\label{eq:secondorder2}
\partial_{t}F^{(1)}&=&\mathcal{L}_{X}^{(1)}F^{(1)}+\mathcal{L}_{X}^{(2)}F^{(0)}
+\partial_{X}\langle \boldsymbol{Z} |\mathds{G}^{(1)} \boldsymbol{Q} \rangle\, F^{(1)}
\nonumber\\&+&\partial_{X}
\frac{\langle \boldsymbol{Z} |\mathds{G}^{(1)} \boldsymbol{V} \rangle 
\langle\boldsymbol{Q}^{\perp}|(\mathcal{L}_{X}^{(1)}\boldsymbol{Q})\rangle}{\lambda}\,F^{(0)}
\nonumber\\
&+&\partial_{X}
\frac{\langle \boldsymbol{Z} |\mathds{G}^{(1)} \boldsymbol{V} \rangle 
\langle\boldsymbol{Q}^{\perp}|\partial_{X}(\mathds{G}^{(1)}\boldsymbol{Q}\,F^{(0)})\rangle}{\lambda}
\nonumber\\
&+&\frac{1}{2}\partial_{X}^{2}\langle\boldsymbol{Z}|\mathds{G}^{(2)}\boldsymbol{Q}\rangle\,F^{(0)}
\end{eqnarray}
We get into
\begin{eqnarray}\label{eq:F1}
\lefteqn{\hspace{-0.4cm}
\partial_{t}F^{(1)}	=\mathcal{L}_{X}^{(1)}F^{(1)}+\mathcal{L}_{X}^{(2)}F^{(0)}
}
\nonumber\\&& \hspace{-0.4cm}
-\partial_X \left(\sum_{i=1}^{2}\jmath^{(i)}\, F^{(i)}\right)+\frac{1}{2}\partial_X^2\left(
S\,F^{(0)}\right)
\end{eqnarray}
where $\jmath^{(1)}$ and $\jmath^{(2)}$ are respectively specified by (\ref{master:j1}), (\ref{master:j2})
and the diffusion coefficient $S$ is defined by equation (\ref{master:dc})
in the main text.

\paragraph{Order $O(\varepsilon^{2})$ accuracy approximation}

Let us now define $F(X,t)=F_0(X,t)+\varepsilon F_1(X,t)$ then summing 
equations \eqref{eq:F0} 
and $\varepsilon$ times \eqref{eq:F1} reconstruct within $O(\varepsilon)$ accuracy 
the Fokker-Planck equation \eqref{eq:final}.

\paragraph{Positivity of the diffusion coefficient}

By construction the matrix $\mathds{G}^{(2)}$ has positive components. Hence
\begin{eqnarray}
\Delta^{(1)}(X)\,>\,0
\nonumber
\end{eqnarray}
because it is the sum of positive addends. To prove that
\begin{eqnarray}
\Delta^{(2)}(X)\,>\,0
\nonumber
\end{eqnarray}
we observe that the two dimensional matrix $\mathds{G}_{1}$ has the form
\begin{eqnarray}
\mathds{G}^{(1)}=\begin{bmatrix}
m_{1} & m_{2}
\\
-m_{3} & m_{1}
\end{bmatrix}
\nonumber
\end{eqnarray}
for $m_{i}\,\geq\,0$ and $i=1,2,3$. Hence
\begin{eqnarray}
\Delta^{(2)}(X)
&=&2\,\frac{\langle \boldsymbol{Z} |\mathds{G}^{(1)}\mathds{J} \boldsymbol{Z} \rangle 
\langle\boldsymbol{Q}^{\perp}|\mathds{G}^{(1)}\mathds{J}^{-1}\boldsymbol{Q}^{\perp}\rangle}{\lambda}
\nonumber\\
&=&2\frac{(\sum_{r=0}^{1}Z_{r}^{2}m_{r})(\sum_{s=0}^{1}(Q^{\perp}_{s})^{2}m_{s})}{|\lambda|}\,\geq\,0
\nonumber
\end{eqnarray}
%

\end{document}